%% file: arxiv_V3_01_05_2025.tex
\documentclass[reprint,
superscriptaddress,
frontmatterverbose, 
nofootinbib,
 amsmath,amssymb,
 aps,
prl,
floatfix
]{revtex4-2}
\usepackage[utf8]{inputenc}
\usepackage{amssymb,amsmath}
\usepackage{graphicx}

\usepackage{microtype}
\usepackage{float}
\usepackage{bbold}
\usepackage[unicode=true,
  colorlinks = true,
            linkcolor = blue,
            urlcolor  = blue,
            citecolor = blue,
            anchorcolor = blue,
    filecolor=blue]{hyperref}
\input{macros}

\usepackage{soul}

\renewcommand{\bJ}{\mathbb{J}}

\usepackage{mathtools}
\DeclarePairedDelimiter\floor{\lfloor}{\rfloor}

\begin{document}

\title{Thermodynamic phases in  first detected return times of quantum many-body systems}

\author{Benjamin Walter}
\affiliation{Department of Mathematics, Imperial College London, London SW7 2AZ, United Kingdom}
\author{Gabriele Perfetto}
\affiliation{Institut f\"ur Theoretische Physik, Universit\"at T\"ubingen, Auf der Morgenstelle 14, 72076 T\"ubingen, Germany}
\author{Andrea Gambassi}
\affiliation{SISSA--International School for Advanced Studies and INFN, via Bonomea 265, 34136 Trieste, Italy}

\begin{abstract}
We study the probability distribution of the first return time to the initial state of a quantum many-body system subject to global projective measurements at stroboscopic times. 
We show that this distribution can be mapped to a continuation of the canonical partition function of a classical spin chain with noninteracting domains at equilibrium, which is entirely characterized by the Loschmidt amplitude of the quantum many-body system. This allows us to conclude that this probability may decay either algebraically or exponentially
at long times, depending on whether the spin chain displays a ferromagnetic or a paramagnetic phase. We illustrate this idea on the example of the return time of $N$ adjacent fermions in a tight-binding model, revealing a rich phase behavior, which can be tuned by scaling the probing time as a function of $N$. The analysis presented here provides an overarching understanding of many-body quantum first-detection problems in terms of equilibrium thermodynamic phases. Our theoretical predictions are in excellent agreement with exact numerical computations. 
\end{abstract}

\maketitle

\textbf{Introduction---} 
Understanding the propagation of quantum information in monitored many-body quantum systems is of central importance for quantum simulators \cite{bloch2012,georgescu2014,browaeys2020, monroe2021,altman2021} and computation \cite{briegel2009,lanyon2013,aolita2015,arute2019,lewis-swan2019}. 
Thus motivated, the impact of monitoring \emph{local} degrees of freedom in many-body quantum dynamics has recently attracted significant interest, leading to the discovery of measurement-induced phase transitions \cite{chan2019,skinner2019a,potter2022,fisher2023}. For quantum computing applications \cite{nielsen2010,childs2013,cheuk2015,nakayama2015,schafer2020,gross2021a}, however,  \emph{global} site-resolved monitoring of the many-body wave function is also fundamental. In spite of this crucial role, the interplay between unitary dynamics and global projective measurements has been rarely investigated so far.

\begin{figure}[b]
\includegraphics[width=\columnwidth]{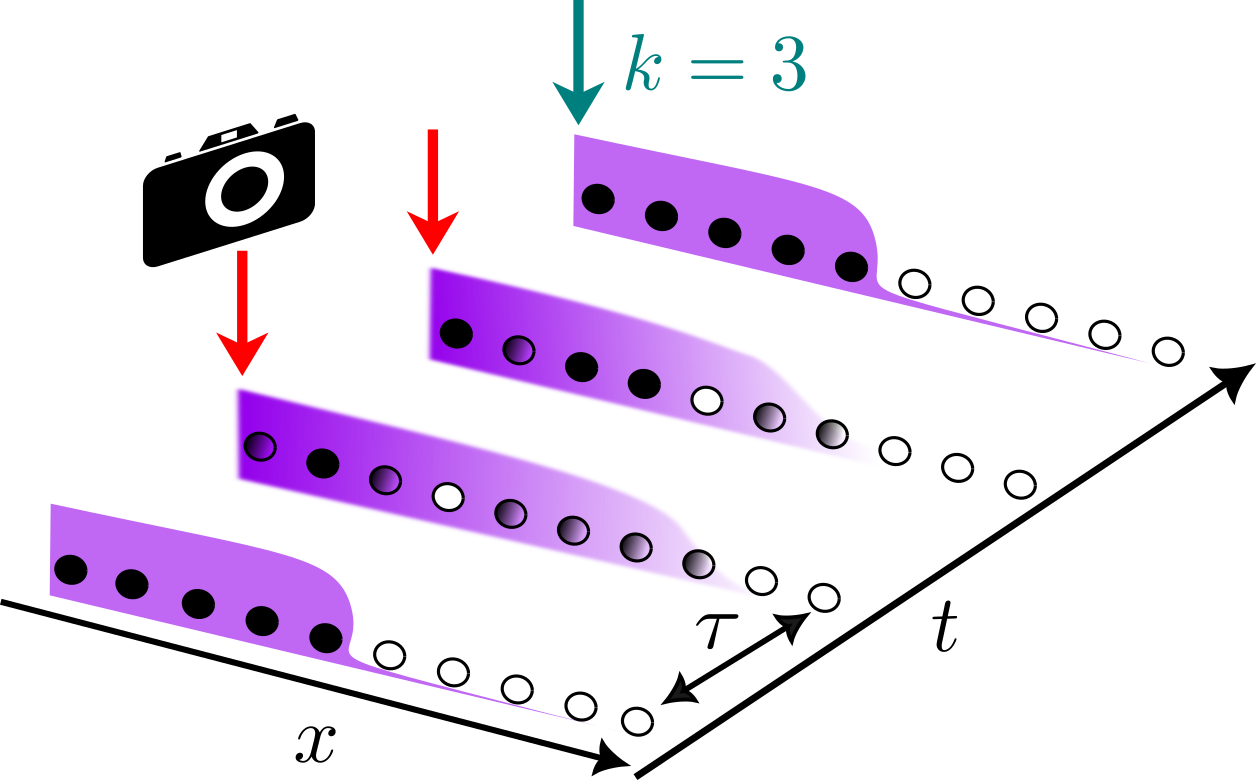}
\caption{\textbf{Illustration of the many-body first-detected return protocol}: A wave function is prepared in an initial state $\ket{\Psi(0)}$ with $N$ adjacent particles and it evolves unitarily; at stroboscopic times $\tau,2\tau, \dots, k\tau$, a projective measurement detects whether or not the state has returned to $\ket{\Psi(0)}$. The first such successful measurement defines the FDRT, here $3 \tau$.}
\label{fig:cartoon}
\end{figure}

In order to advance in this direction, we tackle this problem by studying the \emph{first-detected return time} (FDRT) of quantum many-body systems under global projective measurements at stroboscopic times $\tau,2\tau, \ldots, k\tau $, see Fig.~\ref{fig:cartoon}. This ``stroboscopic'' protocol, studied for single particles in Refs.~\cite{QuantumWalksTodd2006,grunbaum2013recurrence,dhar2015,dhar2015b,Sink2015,friedman2016,friedman2017,thiel2018,thiel2018a,thiel2019,liu2020,darkstates2020,Nonhermitian2020,quantumdyn2021,das2022quantum,yin2023,yin2023a,majumdar2023,chatterjee2023quest,yin2024restart,wang2024first}, has been implemented on a quantum computer \cite{tornow2023,wang2024first,yin2024restart}, 
demonstrating a potential speedup of quantum search algorithms. For quantum many-body systems, however, analytical results have been  obtained  for mean values  only \cite{liu2024properties}. For the first-detection distribution, results are limited to small system sizes \cite{dittel2023,Imparato2024} due to the exponential complexity of many-body numerical simulations. No analytical understanding of many-body FDRT is currently within reach of established techniques.

In this Letter, we investigate the \emph{full} FDRT distribution $F_k$ for arbitrary many-body systems, deriving exact analytical predictions for it. We do so by devising an exact mapping of the quantum many-body FDRT probability onto the classical partition function of a one-dimensional spin model with non-interacting domains. Similar classical partition functions appear in studies of the DNA melting transitions \cite{poland1966,poland1966a,fisher1984,richard2004}, in the truncated inverse distance squared Ising model \cite{bar2014,bar2014a,barma2019,mukamel2023}, or in non-equilibrium currents \cite{harris2017}. In the mapping we devise, the canonical weight of classical spins is obtained from the many-body Loschmidt amplitude. From the latter, we find that the spins within a magnetic domain 
feature long-range interactions and  thus phase transitions already in one dimension. Specifically, the mapping proposed here allows us to classify and understand the different asymptotic decays of $F_k$ as a function of $k$ in terms of the equilibrium phases of the associated classical model. Namely, ferromagnetic ordering is mapped onto algebraic decay of $F_k$, while paramagnetic behavior corresponds to a faster exponential decay of $F_k$. We illustrate this approach by considering the case of a domain of $N$ adjacent fermions in a tight-binding model. Different thermodynamic phases in $F_k$ are then understood as general emergent phenomena due to transport of quasi-particle modes over large-scale distances set by the initial domain length $N$. For free fermions, where transport is ballistic, the onset of the two phases can be tuned by scaling the probing time $\tau$ linearly with $N$. The mapping presented here and the resulting thermodynamic phases are fundamentally different from those discussed for a single qubit in Refs.~\cite{PRAweak1,PRAweak2}, where the long-range interaction emerges only for weak measurements. Accordingly, phase transitions in the measurement output of single qubits cannot be observed for purely projective measurements, in stark contrast with what is predicted here for many-body systems. Our analytical predictions highlight the rich impact of many-body quantum dynamics on FDRT and a deep connection between non-equilibrium measurement-induced quantum fluctuations and equilibrium thermodynamic phases. 
%
\begin{figure}[t]
\includegraphics[width=0.9\columnwidth]{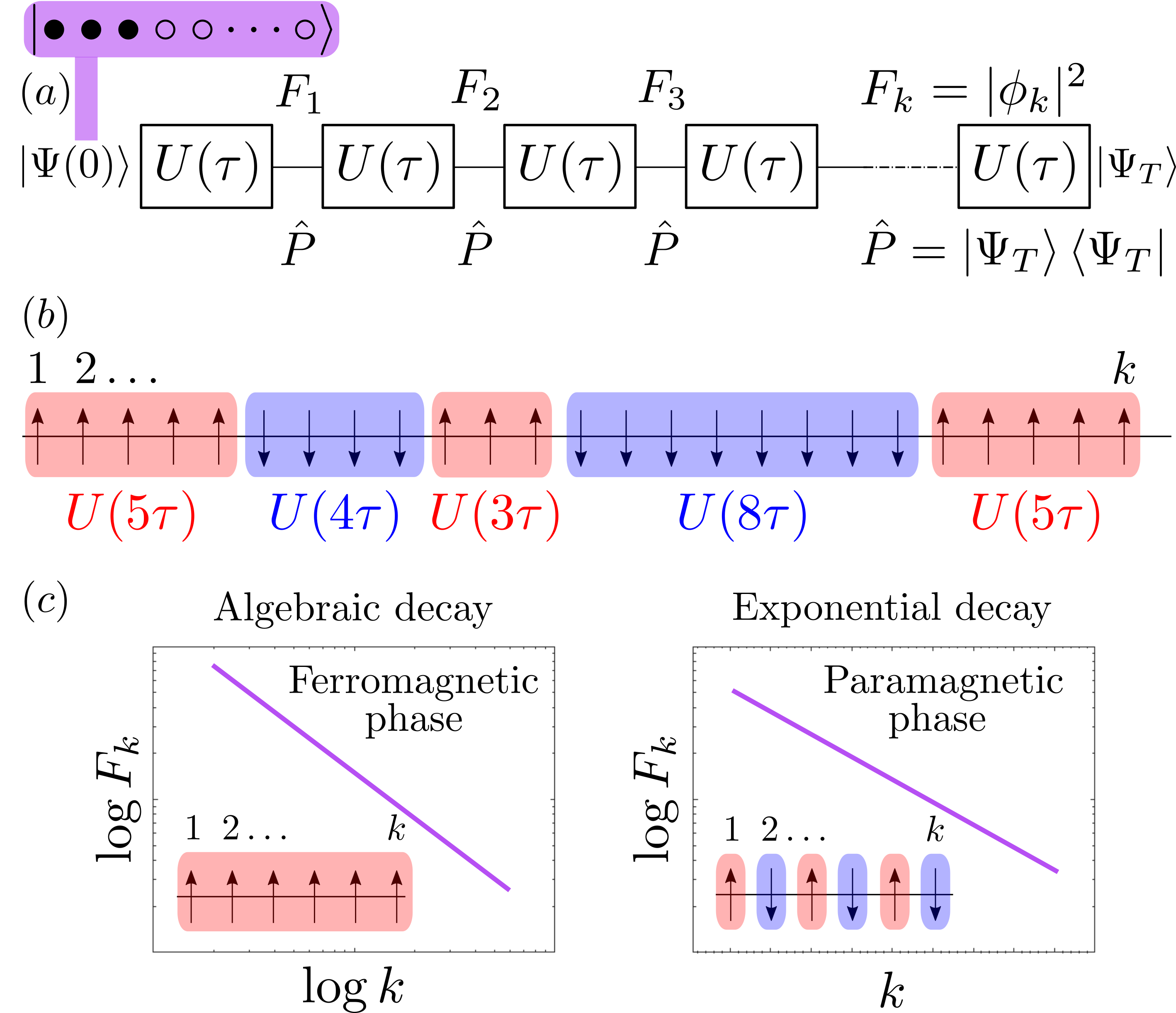}
\centering
    \caption{\textbf{Mapping between quantum many-body first detection time and statistical mechanics.} $(a)$~A wavefunction evolves with unitary dynamics $U(\tau)$ interspersed by projective measurements $\hat P$ occurring at stroboscopic times multiples of $\tau$. The target state $\ket{\Psi_T} = \ket{\Psi(0)}$ is first detected at the $k$th measurement with first detected return amplitude $\phi_k$ and probability $F_k=|\phi_k|^2$. The protocol terminates at the first successful detection in $\ket{\Psi_T}$. 
    $(b)$~$\phi_k$ can be decomposed in terms of time intervals of measurements-free dynamics $U(n\tau)$ (in the figure $n=5,4,3,8,$ and $5$, from left to right). Each interval can be mapped into a domain of aligned classical spins, with spins interacting via long-range interactions. Accordingly, $\phi_k$ is mapped onto a problem of statistical mechanics of domains at equilibrium in a volume $k$. 
    $(c)$~When $F_k$ is determined by long measurement-free intervals $\sim U(k\tau)$, one has a ferromagnetic phase and $F_k$ decays algebraically upon increasing $k$. Conversely, when short intervals $\sim U(\tau)$ dominate, $F_k$ decays exponentially. This corresponds to the paramagnetic phase.}
    \label{fig:spin_model}
\end{figure}

\textbf{Quantum first detected return and statistical mechanics ---}
The FDRT probability of many-body systems described by a Hamiltonian $\cH$, with associated unitary evolution $U(t)=e^{-i \cH t}$, can be tackled by extending the formalism developed in Refs.~\cite{grunbaum2013recurrence,dhar2015, friedman2016,friedman2017} to the many-body realm, see Fig.~\ref{fig:spin_model}$(a)$. Right before the first measurement 
at time $\tau$, the overlap of the wavefunction $\ket{\Psi(\tau^-)} = U(\tau)\ket{\Psi(0)}$ with the target state $\ket{\Psi(0)}$ is given by the Loschmidt amplitude
\begin{align}
\label{eq:loschmidt_def}
    \mathcal{L}(\tau) = \bra{\Psi(0)} U(\tau)
    \ket{\Psi(0)},
\end{align}
and therefore the detection in $\ket{\Psi(0)}$ is successful with probability $F_1 = |\cL(\tau)|^2$. In this case, one has 
$\ket{\Psi(\tau^+)} = \ket{\Psi(0)}$.
If the detection is, instead, unsuccessful, the wavefunction is projected onto the space orthogonal to the target state $\ket{\Psi(\tau^+)} = (\mathbb{1}- \hat{P}) \ket{\Psi(\tau^-)}/\sqrt{1-F_1}$, where 
$\hat{P} = \ket{\Psi(0)}\bra{\Psi(0)}$ such that $\braket{\Psi(\tau^+)}{\Psi(\tau^+)} = 1$. 
Until the next measurement, i.e., for $\tau < t < 2 \tau$   
the state evolves unitarily,
whence the procedure is repeated until the first detection in the target state occurs. The overlap between the target state and the unnormalised wavefunction after $k-1$ unsuccessful detections evolved until immediately before the $k$th measurement defines the \emph{first detection amplitude},
\begin{equation}
   \phi_k = \bra{\Psi(0)} U(\tau) \left[ \left( \mathbb{1}  - \hat{P}\right) U(\tau)\right]^{k-1} \ket{\Psi(0)}.
    \label{eq:def_phi}
\end{equation}
The FDRT probability $F_k$ is given by $F_k = |\phi_k|^2$. 
The generating function
$\hat{\phi}(z) = \sum_{k \geq 1} \phi_k z^k$ of $\phi_k$ is related to that $\hat{\cL}(z) = \sum_{k \geq 1} \cL_k z^k$ of $\cL_k \equiv \cL(k \tau)$ by \cite{grunbaum2013recurrence,dhar2015b,friedman2016}
\begin{align}
    \hat{\phi}(z) = \frac{\hat{\cL}(z)}{ 1+ \hat{\cL}(z)}.
\label{eq:phihat_as_Lhat_ratio}
\end{align}
The $\phi_k$'s are then determined  as the integral
\begin{equation}
\label{eq:inv-phik}
    \phi_k = \oint \frac{\dint z}{2\pi i} \frac{\hat{\phi}(z)}{z^{k+1}},
\end{equation}
along a contour enclosing the origin $z=0$ and within 
the domain of analyticity of $\hat{\phi}(z)$. 

By expanding the $(k-1)$th power of $(\mathbb{1}-\hat{P})U(\tau)$ in 
Eq.~\eqref{eq:def_phi}, one finds that $\phi_k$ can be written as
\begin{align}
    \phi_k = \sum_{r = 1}^k (-1)^{r+1} \!\!\!\!\!\sum_{\ell_1, \ell_2, \ldots,\ell_r = 1}^{\infty} 
     \left( \prod_{j=1}^r \cL_{\ell_j} \right) \delta_{\sum_{j=1}^r \ell_j,k}.
\label{eq:phi_explicit}
\end{align}
Equation \eqref{eq:phi_explicit} is crucial to map the problem into the partition function of a classical lattice spin model with non-interacting domains, as sketched in Fig.~\ref{fig:spin_model}$(b)$. A configuration of the spin model corresponds to a partition of the total volume $k$ into $r$ consecutive domains of lengths $\ell_1, ..., \ell_r$, with $\ell_1 + ... + \ell_r = k$ (see Fig.~\ref{fig:spin_model}). Each domain has an associated length-dependent energy $E(\ell)$ resulting from the possibly long-range interactions among the spins within the domain. The ``Boltzmann weight'' $w$ is associated, instead, to each domain wall.
The canonical partition function of this model with 
fixed volume $k$ can be written as
\begin{align}
    Z(k,w) = \sum_{r = 1}^k w^{r+1} 
    \!\!\!\!\!\sum_{\ell_1, \ell_2, \ldots,\ell_r = 1}^{\infty} 
    \left( \prod_{j=1}^r e^{-E(\ell_j)} \right)\delta_{\sum_{j=1}^r \ell_j,k}.
        \label{eq:canonical_partition}
\end{align}
It is then apparent that \Eqref{eq:phi_explicit} can be recovered from \Eqref{eq:canonical_partition} by letting $w\to -1$ and by identifying the Loschmidt amplitude $\cL_\ell= \cL(\ell \tau)$ 
in time 
with the Boltzmann factor $e^{-E(\ell)}$ in space. 
Differently from the classical model, $E(\ell) = - \ln \cL(\ell \tau)$ may now take complex values depending on the sign of $\cL_{\ell}$, while $w=-1$ implies a purely imaginary energetic cost associated to a domain wall; these differences give rise to interferences when summed over, reflecting the quantum nature of $\phi_k$. The analogy extends to the fixed-pressure grand canonical ensemble. Summing over the length $k$, one obtains the partition function
\begin{align}
    \mathcal{Z}(z,w) &= 
    \sum_{k=1}^{\infty} z^k Z(k,w) = \frac{w^2 \hat{\cL}(z)}{1 - w \hat{\cL}(z)},
\label{eq:grand_canonical_partition}
\end{align}
where $z$ is 
the exponential of the pressure conjugate to the volume $k$, while $\hat{\cL}(z) = \sum_{k \geq 1}  e^{-E(k)} z^k$ is the grand canonical partition function of a single domain. Setting $w\to -1$ in Eq.~\eqref{eq:grand_canonical_partition}, one recovers $\hat{\phi}(z)$  
given in \Eqref{eq:phihat_as_Lhat_ratio}. 

\textbf{Emergence of thermodynamic phases---} 
In the thermodynamic limit $k\to \infty$, the spin model introduced in Eq.~\eqref{eq:canonical_partition} may exhibit, depending on the form of $E(\ell)$ and on the value of $w$, either a ferromagnetic or a paramagnetic phase \cite{bar2014,bar2014a,barma2019,mukamel2023}. This may happen even in one spatial dimension since  $E(\ell)$ results from possibly long-range interactions among the spins within the domain. In the former case, the equilibrium ensemble is dominated by a globally ordered spin domain of length $\ell \sim \cO(k)$, while in the latter by disordered spins with $\ell \sim \cO(1)$. Identifying $\hat{\phi}(z)$ as $\cZ(z,w \to -1)$
allows us to establish the possibility of observing two different behaviors of $\phi_k$ at large $k$, as shown in Fig.~\ref{fig:spin_model}$(c)$.
If the spin model is in  the ferromagnetic phase, the dominant configuration is a single, extended, domain and thus $Z(k,-1) \sim e^{-E(k)}$; according to the mapping, this corresponds to the leading term of the first-detection amplitude  being  $\phi_k \sim \cL(k \tau)$.
If the spin model is, instead, in the paramagnetic phase, and therefore it displays disordered domains, 
the partition function is dominated by many short domains $Z(k,-1)\sim e^{-kE(1)}$. In terms of the quantum evolution, this corresponds to $\phi_k \sim \left[ \cL(\tau) \right]^k$. 

The two cases mentioned above are understood through the analytic properties of $\hat{\phi}(z) = \cZ(z,-1)$ as a function of $z$ which determines $\phi_k$ according to Eq.~\eqref{eq:inv-phik}. The radius of convergence of $\hat{\phi}$ is controlled by its closest singularity $z^*$ to the origin. By inspection of Eqs.~\eqref{eq:phihat_as_Lhat_ratio} and \eqref{eq:grand_canonical_partition}, the singular point $z^*$ may arise from either $(i)$ a root in the denominator, i.e., $ 1+ \hat{\cL}(z^*) = 0$, leading to  a simple pole, or $(ii)$ a non-analyticity in $\hat{\cL}(z^*)$ leading to a branch point. If the singularity at $z^*$ is a simple pole, $\hat{\phi}(z) \sim (z-z^*)^{-1}$, the integration in Eq.\eqref{eq:inv-phik} renders  $\phi_k \sim (z^*)^{-k}$ and therefore an exponential dependence of $\phi_k$ on $k$. This exponential decay corresponds to the configuration with $k$ domains, i.e., to the paramagnetic phase. If the singularity $z^*$ is a branch point, instead, $\hat{\phi}(z) \sim |z-z^*|^{c-1}$ and, via Tauberian theorems, one obtains from \eqref{eq:inv-phik} an algebraic decay of $\phi_k \sim k^{-c}$ for large $k$ \cite{flajolet2009}. Algebraic growth of a canonical partition function as a function of the volume is a consequence of a single domain in a long-range interacting spin model and thus we associate this behavior with the ferromagnetic phase. 

We illustrate this mapping by investigating the FDRT of $N$ free fermions with Hamiltonian $\cH =  \sum_{j=-L}^{L} \left( c_j^{\dagger} c_{j+1} +c_{j+1}^{\dagger} c_{j} \right)$,
where $c_j$ ($c^\dagger_j$) is the fermionic annihilation (creation) operator at lattice site $j$. The system size $L$ is henceforth taken to infinity $L\to \infty$. We consider an initial state $\ket{\Psi(0)}$ consisting of $N$ adjacent fermions $\ket{\Psi(0)} = \otimes_{j=1}^N c_j^{\adj} \ket{0}$, where  $\ket{0}$ is the vacuum of the system [see the sketch in Fig.~\ref{fig:cartoon} and in Fig.~\ref{fig:spin_model}$(a)$]. The case $N=\infty$ corresponds to a domain wall at site $j=0$ separating the empty half-infinite chain from the complementary filled one \cite{viti2016,wei2017}. The free fermion model is motivated by recent implementation of quantum walks under stroboscopic measurements on quantum computers \cite{tornow2023,wang2024first,yin2024restart}, where a qubit hopping Hamiltonian analogous to $\mathcal{H}$ above, with 
$N=1$, is implemented. 

\textbf{Ferromagnetic phase for finite $N$---} 
For finite $N$, $\cL_k$ is  the determinant of the matrix  $\bJ_N( k \tau) = \left[ i^{n-m}J_{m-n}(2k  \tau) \right]_{m,n = 1 ... N}$, where $J_{n}(x)$ is 
the $n$th Bessel function of the first kind \cite{krapivsky2018}. For large times $k \tau$, the determinant $\cL_k$ decays algebraically \cite{krapivsky2018,SM-ref}. 
According to the mapping discussed above, the energy cost $E(\ell) = - \ln \cL(\ell \tau)$ of a domain is logarithmic in $\ell$
\begin{align}
\label{eq:energy_ell}
 E(\ell) = 
     c_N \ln \ell + \Delta_N + (N \, {\rm mod} \, 2) \ln \cos\left( 2  \ell \tau - N \pi/4 \right),
\end{align}
where we introduced
$c_N = (N^2+ N \, {\rm mod} \, 2)/4$. The definition of $\Delta_N$ is reported in the Supplemental Material \cite{SM-ref, barnes1901}. 
Introducing the fugacity $y =  e^{-\Delta_N}$,  the grand canonical partition function of the spin model in \Eqref{eq:grand_canonical_partition} reads
\begin{align}
    \cZ_N(z,w) =
        \frac{ w^2 y \Lambda_N(z)}{1 - w y \Lambda_N(z)},
\label{eq:Z_N}
\end{align}
where $\Lambda_N(z)$ is determined in terms of the polylogarithm function \cite{SM-ref}. For $w = 1$ and $N$ even, Eq.~\eqref{eq:Z_N} equals the partition function  of the truncated inverse  distance squared Ising model (TIDSI) \cite{bar2014} which is parametrized by $y$ and $c_N$. In TIDSI model, spins within the same domain experience a long-range interaction $\sim r^{-2}$ as a function of their mutual distance $r$. In the quantum problem, these long-range interactions 
allow the emergence of 
a ferromagnetic phase for purely projective measurements. Namely, the TIDSI model undergoes a mixed-order phase transition from a paramagnetic to a ferromagnetic phase \cite{bar2014,bar2014a,barma2019,mukamel2023}. In the quantum picture, however, we consider the partition function $\cZ(z,w)$ at $w = -1$. For the spin model, this corresponds to studying the TIDSI model at the (classically forbidden)  negative fugacity $-y$.  

\begin{figure}[t]
 \includegraphics[width=1\columnwidth]{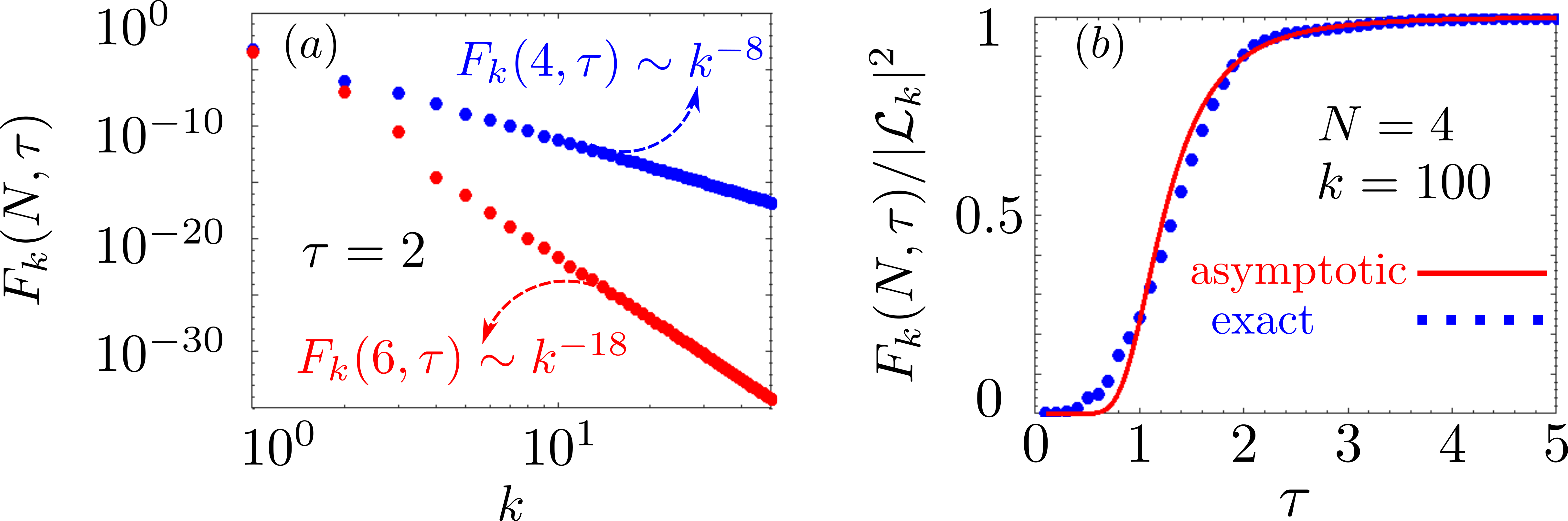}
\caption{\textbf{Ferromagnetic phase - algebraic decay}. $(a)$~Log-log plot of the first detected return time probability $F_k$ as a function of $k$ for fixed stroboscopic time $\tau=2$, with $N=4$ and $6$. $F_k \sim |\cL_k|^2 \sim k^{-2c_N}$ (with $c_4=4$ and $c_6=9$) 
decays algebraically upon increasing $k$ according to the mapping to the ferromagnetic phase of the spin model. 
$(b)$~Ratio $F_k/|\cL_k|^2$ (blue dots) as a function of $\tau$, for fixed $k=100$ and
$N=4$ compared with the branch-cut asymptotic (red solid line). 
} 
\label{fig:ferromagnetic_phase}
\end{figure}

The leading singularity of $\cZ_N$ in Eq.~\eqref{eq:Z_N} is due to branch points \cite{SM-ref, bateman1953,NIST:DLMF}. The integral in Eq.~\eqref{eq:inv-phik} can then be worked out for all $N$, from which one obtains the leading large-$k$ behavior of $\phi_k$. For $N=1$, our analysis renders the result $F_k\sim k^{-3}$ of Ref.~\cite{friedman2017}. For $N > 1$, instead, it gives, up to some oscillating prefactor, $\phi_k \sim \cL_k$ and thus the algebraic decay 
\begin{equation}
\phi_k \sim k^{-c_N} \quad\mbox{and}\quad F_k \sim k^{-2 c_N},
\label{eq:ferromagnet_main_result}
\end{equation}
with logarithmic corrections $\phi_k \sim (\ln k)^n/k^{c_N}$ of all integer orders $n\geq 1$ for $N=2$ \cite{SM-ref}. This algebraic behavior is consistent with a ferromagnetic, globally ordered phase in the spin model. One can intuitively understand this as a consequence of the logarithmically slow growth of the domain energy $E(\ell)$ in length $\ell$, which renders a global ordering thermodynamically favourable in the large $k$ limit and therefore $\phi_k \sim \cL_k$. In Fig.~\ref{fig:ferromagnetic_phase}$(a)$, we show the exact FDRT probabilities $F_k$ for  $N=4$ and $6$ \cite{SM-ref}, with fixed $\tau=2$, as obtained from the exactly known Loschmidt echo, which are compared with the expected leading algebraic behavior \eqref{eq:ferromagnet_main_result}, showing excellent agreement for large $k$. Figure~\ref{fig:ferromagnetic_phase}$(b)$ shows that the ratio 
$F_k/|\cL_k|^2$ is well captured by the branch cut asymptotics for intermediate values of $\tau \gtrsim 1.8$ and large values of $k$.

\begin{figure}[t]
    \centering    \includegraphics[width=1\columnwidth]{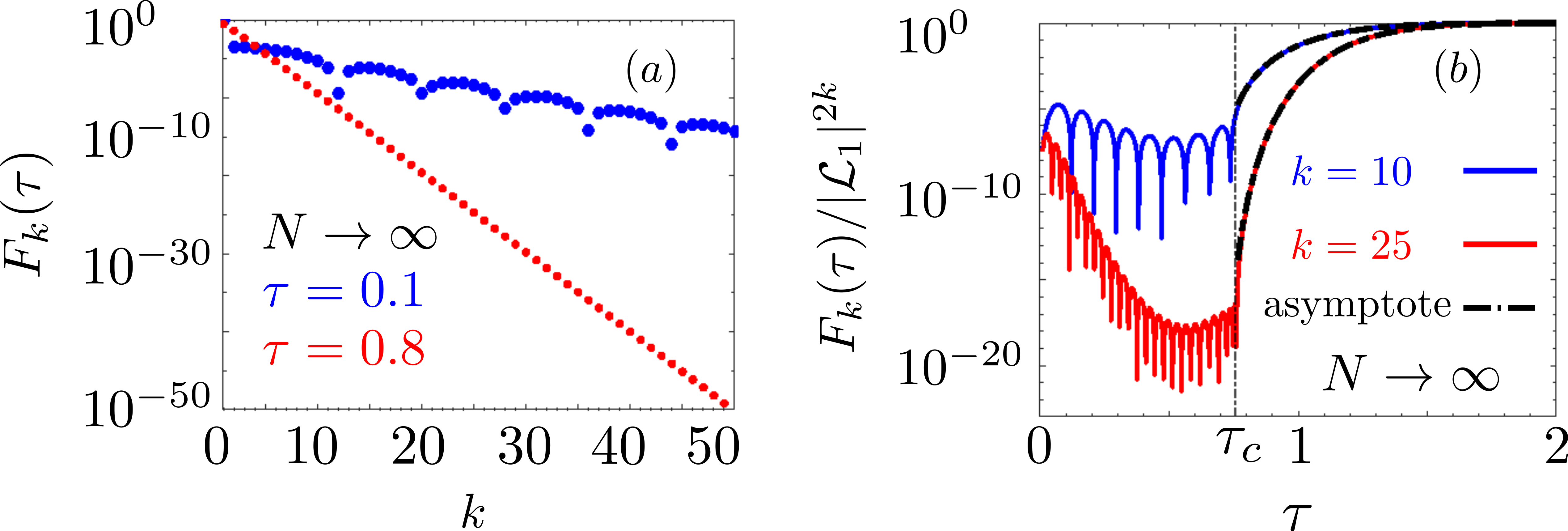}     \caption{\textbf{Paramagnetic phase - exponential decay}. $(a)$~Log-linear plot of the first detected return time probability $F_k(\tau) \equiv F_k(N = \infty,\tau)$ of the semi-infinite occupied line as a function of $k$ for stroboscopic times $\tau = 0.1$ (top-blue dotted) and $0.8$ (bottom-red dashed).
     $F_k$ decays exponentially upon increasing $k$, according to the mapping to the paramagnetic phase of the spin model. This exponential decay holds for all values of $\tau$.
     For $\tau < \tau_c = 0.77\ldots$, $F_k$ features additional oscillations (blue dots). 
     $(b)$~Log-linear plot of the ratio $F_k(\tau)/|\cL_1|^{2k}$ as a function of $\tau$ for fixed values of $k=10$ and 25. This ratio, for $\tau > \tau_c$, is well approximated by the systematic expansion (dot-dashed black lines) mentioned in the main text. 
     For $\tau < \tau_c$, instead, the asymptotics break down  and $F_k(\tau)$ displays $k-2$ positive roots as a function of $\tau$ .
     } 
     \label{fig:N_inf}
\end{figure}

\textbf{Paramagnetic phase for $N = \infty$---}
For an half-infinite domain wall, i.e., for $N=\infty$, the Loschmidt amplitude is exactly given by  $\cL(\tau) = e^{-\tau^2}$ \cite{viti2016,wei2017}.
In the spin model this corresponds to a quadratic energetic cost of a domain: $E(\ell) = \tau^2 \ell^2$. In order to evaluate $\cZ_N(z,w)$ in Eq.~\eqref{eq:grand_canonical_partition}, we introduce a partial theta function \cite{Andrews1981}
    $\Theta_+(z,q) = \sum_{k = 1}^{\infty} z^k q^{k^2}$
in terms of which $\hat{\cL}(z) = \Theta_+(z,e^{-\tau^2})$. The grand canonical partition function is  
\begin{align}   \mathcal{Z}_{\infty}(z,w) = \frac{w^2 \Theta_+(z,e^{-\tau^2})}{1-w \Theta_+(z,e^{-\tau^2})}.
   \label{Z_para}
\end{align}
For all $q \in (0,1)$, $\Theta_+(z,q)$ is an analytic function of $z$ \cite{Kostov2023}, thus excluding branch cuts. In turn,  for $w=-1$, $\mathcal{Z}_{\infty}$ has a countable set of poles at the complex roots $z_n$ of $1 + \Theta_+(z_n,q) = 0$ \cite{hardy1905}. 
For $q^{-2} \geq q_{\infty} = 3.23\ldots$, these roots are simple, real and negative \cite{katkova2004,kostov2013}, and they are approximately given, for large $n$, by $z_n=-q^{1-2 n}$ \cite{hutchinson1923}. This implies that for $\tau \geq \tau_c =  \sqrt{(\ln q_{\infty})/2} = 0.77\ldots$ the contour integral in Eq.~\eqref{eq:inv-phik} is dominated, for large $k$, by the simple pole $z^* = z_1$ closest to the origin.
One then finds $\phi_k \sim (z^*)^{-k}$  for $\tau > \tau_c$ with a $k$-independent prefactor. This exponential decay is shown in Fig.~\ref{fig:N_inf}$(a)$ for $\tau=0.8>\tau_c$. 
This behavior corresponds to the paramagnetic phase in the spin model and can therefore  be explained with the stronger quadratic growth of $E(\ell)$ upon increasing $\ell$, compared to the logarithmic growth \eqref{eq:energy_ell} for finite $N$. For $\tau<\tau_c$, the poles of $\cZ_{\infty}$ are complex conjugate and therefore $F_k$ displays oscillations as a function of $k$. This is illustrated in Fig.~\ref{fig:N_inf}$(a)$ for $\tau = 0.1 < \tau_c$. The critical value $\tau_c$ is not related to the energy bandwidth of the model and it is therefore a fundamentally novel feature compared to the ``resonant probing times'' \cite{friedman2017,SM-ref} occurring at finite $N$. The asymptotics of $F_k/|\cL_1|^{2k}$ for $\tau>\tau_c$ can be obtained by applying Fa\`{a} di Bruno's theorem \cite{gould1972,comtet1974} to $\mathcal{Z}_{\infty}$, which amounts to 
performing an expansion in the maximal domain length $\ell_{\rm M}$ considered \cite{SM-ref,gould1972,comtet1974}. 
The estimates with $\ell_{\rm M} = 3$ 
are compared in Fig.~\ref{fig:N_inf}$(b)$ with the exact prediction for $F_k/|\cL_1|^{2k}$, showing excellent agreement. As $\tau \to 0$ the quantum Zeno regime is met. Surprisingly, we observe \cite{SM-ref} that, correspondingly, $F_k(N,\tau)$ coincides with $F_k(N\to\infty,\tau)$ for every value of the particle number $N$.

\begin{figure}  
\includegraphics[width=1\columnwidth]{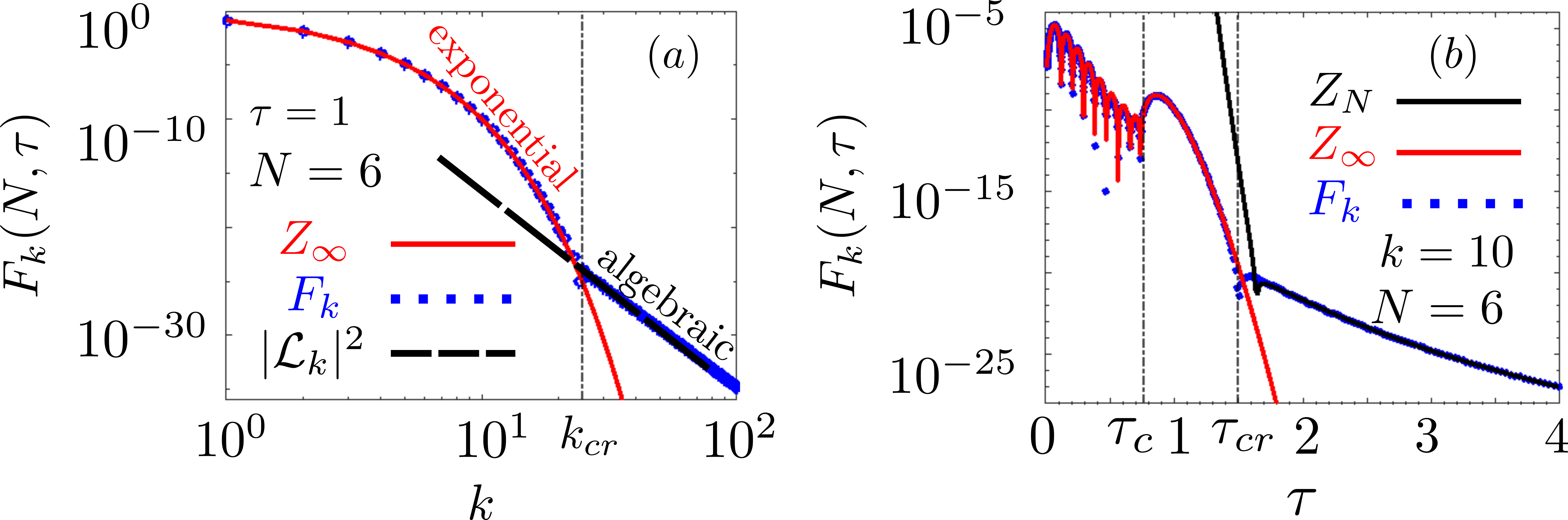}
             \caption{\textbf{Phase crossover.} $(a)$~Log-log plot of the first detection probability $F_k(N,\tau)$ as a function of $k$ for a fixed $\tau=1$ and $N=6$. 
             Upon increasing $k$, $F_k$ shows a crossover from an exponential-paramagnetic to an algebraic-ferromagnetic decay. In this panel, the crossover occurs at $k = k_{cr} \approx 25$ for $N=6$, where $k_{cr}\sim N/\tau$. 
             $(b)$~Log-linear plot of $F_k(N,\tau)$ as a function of $\tau$ for fixed $k=10$ and for $N=6$ fermions. For $\tau < \tau_{cr}$, with $\tau_{cr}\approx 1.5$ and $\tau_{cr} \sim N/k$, $F_k$ (blue dots) coincides with the paramagnetic partition function $|Z_{\infty}(k,-1)|^2$ (red line), while for $\tau > \tau_{cr}$, $F_{k}(\tau)$ coincides, up to some proportionality factor, with the ferromagnetic partition function $|Z_N(k,-1)|^2$ (black line). The crossover stroboscopic time scales as $\tau_{cr} \sim N/k$ upon increasing $N$.
             }
     \label{fig:crossover}
\end{figure}

\textbf{Phase crossover and ferromagnetic transition---}
We investigate here the onset of the ferromagnetic phase in $F_k$ upon increasing $k$ beyond a crossover volume $k_{cr}$. 
For sufficiently small $k<k_{cr}$, 
$F_k$ is in perfect agreement with the 
paramagnetic behavior observed for $N=\infty$ fermions and predicted by $Z_{\infty}(k,-1)$ in Eq.~\eqref{Z_para}.
Upon increasing $k$ beyond $k_{cr}$, instead, $F_k$ sharply crosses over into the ferromagnetic behavior predicted by Eq.~\eqref{eq:Z_N}, where $F_k \sim |\cL_k|^2$. This crossover is illustrated in Fig.~\ref{fig:crossover}$(a)$ for $N=6$ and $\tau=1$. Note that, upon increasing $\tau$, the initial exponential decay might not be visible, as it happens in Fig.~\ref{fig:ferromagnetic_phase}$(a)$ with $\tau=2$, where only the eventual algebraic behavior is seen. In a complementary way, Fig.~\ref{fig:crossover}$(b)$ shows that for a fixed value of $k=10$, larger values of $\tau>\tau_{cr}$ are needed in order for $F_k$ to be described by the ferromagnetic partition function in Eq.~\eqref{eq:Z_N}.     

For large $N$, this crossover from paramagnetic to ferromagnetic behavior can be rationalized by studying the scaling limit of the Loschmidt amplitude $\cL(\tau)$. Keeping  $x = \tau/N$ finite, $\cL(\tau)$  satisfies $\lim_{N \to \infty}  N^{-2} \ln \cL(N x) = -f(x)$ where, up to oscillating factors for $N$ odd, $f(x) \approx (\ln x)/4 + 3/8$ for $x \gg 1$ and $f(x) \approx x^2$ for $x \ll 1$ \cite{krapivsky2018}. This fact implies that, upon rescaling the stroboscopic time $\tau$ with the number $N$ of fermions, the energy $E(k) = N^2 f(k \tau/N)$ associated to a domain in the spin model may either grow logarithmically or quadratically as a function of $k$.  
We conclude that the typical crossover chain volume $(k\tau)_{cr}$, above which $F_k$ decays algebraically, satisfies $ (k \tau)_{cr} \sim N$ \cite{SM-ref}. We emphasize that crossover from paramagnetic to ferromagnetic decay is a general phenomenon caused by the transport of quasi-particles over the macroscopic distance $N$ set by the initial domain wall state. The only specific aspect to free fermions is provided by the linear scaling between space and time due to the associated ballistic transport \cite{freehd0,freehd2,freehd3,freehd4,viti2016,freehd6,freehd7,freehd8,freehd9,freehd10,freehd11}. 

\textbf{Summary and outlook---} We derived exact analytical predictions for the first detected return time (FDRT) of many-body quantum systems by mapping the FDRT onto the partition function of a classical system of magnetic domains in one spatial dimension. This mapping provides a physical explanation of the rich behavior of the probability of the FDRT which may display an algebraic (ferromagnetic) or an exponential (paramagnetic) decay at long times. Our results are valid for an arbitrarily large number $N$ of particles, including the limit $N\to \infty$. The onset of the algebraic decay can be remarkably controlled by scaling the stroboscopic time $\tau$ with the number $N$ of particles, making such a decay accessible already for small values of $k$, when $F_k$ is still significant. This fact 
is crucial for actual implementations of the protocol discussed here on quantum computer platforms as in Refs.~\cite{tornow2023,wang2024first,yin2024restart}. Therein, indeed, $F_k$ at small values of $k$ can be measured with great accuracy for a small number $N=2$, 3, 4 of particles and rings of finite length $L$. The present analysis therefore paves the way for investigations of collective effects in multiparticle quantum walks via quantum computers. Upon increasing $k$, the probability $F_k$ becomes rapidly small since the initial inhomogeneous domain wall state evolves towards an homogeneous density profile \cite{tasaki2024heat}. In order to devise efficient detection protocols, it is then very interesting to introduce resetting \cite{yin2023,qreset0,qreset1,qreset2,qreset3,qreset4,qreset5,qreset6,majumdar2023,chatterjee2023quest} after a sequence of unsuccessful detections. Similarly, a very promising avenue is provided by models featuring Wannier-Stark localization \cite{WannierWS,FogedbyWS,hartmannWS} or quantum many-body scars \cite{qscars1,qscars2,qscars3,qscars4,qscars5}. In both the cases, the Loschmidt echo shows perfect revivals in time. In the former, this happens for generic initial and target states due to Bloch oscillations induced the confining field. In the latter, revivals are present only when the initial and target state of detection belong to the scarred subspace, since the corresponding energy levels are equispaced. These revivals can then be exploited in order to obtain an enhanced FDRT sensing for large times $k$. In addition, it would be interesting to extend the many-body first detection protocol to weak measurements, which have been so far studied \cite{PRAweak1,PRAweak2} only for single qubit systems.   

\textbf{Acknowledgements.---}
The authors would like to thank Eli Barkai for fruitful discussions and helpful comments.
BW would like to thank Giorgio Li, Adam Nahum, and Kay Wiese for insightful discussions. BW and AG acknowledge support from MIUR PRIN project ``Coarse-grained description for non-equilibrium systems and transport phenomena (CO-NEST)” No.~201798CZL. BW acknowledges funding from the Imperial College Borland Research Fellowship. GP acknowledges support from the Alexander von Humboldt foundation through a Humboldt research fellowship for postdoctoral researchers. AG acknowledges support of PNRR MUR project No. PE0000023-NQSTI.

\textbf{Data availability statement.---}
The code and the data that support the
findings of this Letter are available on Zenodo \cite{zenododata}.
\bibliographystyle{apsrev4-2}
\bibliography{first_detection_biblio}

\setcounter{equation}{0}
\setcounter{figure}{0}
\setcounter{table}{0}
\renewcommand{\theequation}{S\arabic{equation}}
\renewcommand{\thefigure}{S\arabic{figure}}

\makeatletter
\renewcommand{\theequation}{S\arabic{figure}}
\renewcommand{\thefigure}{S\arabic{figure}}

\onecolumngrid
\newpage

\setcounter{page}{1}

\setcounter{secnumdepth}{3}
\pagestyle{plain}

\begin{center}
{\Large SUPPLEMENTAL MATERIAL}
\end{center}
\begin{center}
\vspace{0.8cm}
{\Large Thermodynamic phases in first detected return times of quantum many-body systems}
\end{center}
\begin{center}
Benjamin Walter$^{1}$, Gabriele Perfetto$^{2}$, and Andrea Gambassi$^{3}$
\end{center}
\begin{center}
$^1${\em Department of Mathematics, Imperial College London, London SW7 2AZ, United Kingdom}\\
$^2${\em Institut f\"ur Theoretische Physik, Universit\"at T\"ubingen, Auf der Morgenstelle 14, 72076 T\"ubingen, Germany}\\
$^3${SISSA--International School for Advanced Studies and INFN, via Bonomea 265, 34136 Trieste, Italy}
\end{center}

\setcounter{equation}{0}
\setcounter{figure}{0}
\setcounter{table}{0}
\setcounter{page}{1}
\makeatletter
\renewcommand{\theequation}{S\arabic{equation}}
\renewcommand{\thefigure}{S\arabic{figure}}

\makeatletter
\renewcommand{\theequation}{S\arabic{equation}}
\renewcommand{\thefigure}{S\arabic{figure}}

\renewcommand{\bibnumfmt}[1]{[S#1]}
\renewcommand{\citenumfont}[1]{S#1}

\onecolumngrid

\setcounter{secnumdepth}{3}

\noindent In this Supplemental Material we provide the details of the calculations and of the results presented in the main text. In particular, in Sec.~\ref{app_sec_1_exact}, we recall the exact expression of the Loschmidt amplitude of $N$ adjacent fermions on a lattice and of the first detection amplitude $\phi_k$. These expressions have been used in the main text to produce the exact numerical data. 
In Sec.~\ref{sec_app2_Qzeno} we discuss the first detection probability $F_k$ in the quantum Zeno regime of very small probing time $\tau \to 0$. In Sec.~\ref{app_sec_3_asymptotic} we derive, via an integration in the complex plane, the asymptotic expression in Eq.~(10) of the main text for $\phi_k$ and $F_k$ at large $k$. In Sec.~\ref{sec:effective_exp}, we discuss how the onset of the algebraic behavior in Eq.~(10) can be tuned to small values of $k$ upon increasing the probing time $\tau$. In Sec.~\ref{app_sec_4_Bruno}, we eventually detail the derivation of the asymptotic behavior of $F_k(\tau)$ for $N\to \infty$ and $\tau>\tau_c$ via the Fa\`{a} di Bruno formula.

\section{First-detection amplitude of $N$ adjacent free fermions: exact numerical results}
\label{app_sec_1_exact}

We consider a system of free fermions evolving on a chain composed by $L$ lattice sites according to the Hamiltonian
\begin{equation}
\cH =  \gamma \sum_{j=-L/2}^{L/2-1} \left( c_j^{\dagger} c_{j+1} +c_{j+1}^{\dagger} c_{j} \right),
\label{eq:free_fermions_Hamiltonian}
\end{equation}
where the operators $c_j$ and $c_{j}^{\dagger}$ satisfy  the canonical anticommutation relations $\{c_j,c_{j'}^{\dagger}\}=\delta_{j,j'}$. 
Here and in the main text we set the hopping amplitude $\gamma=1$ (as it just provides the timescale of the evolution). 
We also assume periodic boundary conditions along the chain, i.e., $c_{j+L}=c_j$ and we focus on the limit $L\to \infty$ of an infinite chain. 
The initial state of the dynamics $\ket{\Psi(0)}=\otimes_{j=1}^N c_j^{\adj}\ket{0}$, where $\ket{0}$ is the empty lattice, consists of $N$ adjacent fermions. 
A central quantity for the analysis presented in the main text is the Loschmidt amplitude $\cL(\tau)=\bra{\Psi(0)} U(\tau)\ket{\Psi(0)}$, which gives the probability amplitude that the $N$ fermions get back to the initial state $\ket{\Psi(0)}$ after time $\tau$. In this case, the Loschmidt amplitude can be exactly computed for generic $N$, as detailed in Refs.~\onlinecite{krapivsky2018,viti2016}. 
We report here only the result:
\begin{align}
\label{appeq:loschmidt_exact}
    \cL(\tau) = \begin{cases}
        \det \left[ i^{m-n}J_{n-m}(2 \tau) \right]_{m,n=1}^N  & \mbox{for}\quad N < \infty, \\
        e^{-\tau^2} & \mbox{for}\quad  N= \infty,
    \end{cases}
\end{align}
where $J_n(t)$ is the $n$th Bessel function of the first kind. We emphasize that Eq.~\eqref{appeq:loschmidt_exact} is valid in the limit $L\to\infty$, where the Bessel function $J_{n-m}(\tau)$ emerges as the amplitude of finding a fermion, initially at site $m$, at site $n$ at time $\tau$. The limiting case $N=\infty$ represents an initial domain-wall state, where half of the chain is initially filled, while the other half is empty. This latter case is discussed in details in Refs.~\onlinecite{viti2016,wei2017}. For determining the first detection probability (c.f.~Sec.~\ref{app_sec_3_asymptotic}), the asymptotic of the Loschmidt amplitude $\cL(\tau)$ for large $\tau\gg 1$ and generic, finite, $N$ will be useful. 
We report here the corresponding expression, taken from Ref.~\onlinecite{krapivsky2018}:
\begin{align}
\label{appeq:loschmidt_asymptotic}
\cL(\tau) \simeq \begin{cases}
    C_N \, \tau^{-c_N} & \text{for even }N ,\\
     C_N  \cos\left(2 \tau - N \pi/4 \right)\,\tau^{-c_N} & \text{for odd }N.
\end{cases}
\end{align}
We note that, for even $N$, the Loschmidt amplitude decays monotonically and in an algebraic way for large $\tau \gg 1$, while oscillations are superimposed to the algebraic decay for odd $N$. In Eq.~\eqref{appeq:loschmidt_asymptotic}, we denote by 
\begin{equation}
c_N= (N^2+N\mbox{\,mod\,}2)/4
\label{eq:app-cN}
\end{equation}
the exponent of the algebraic decay, while the amplitude $C_N$ is given by 
$C_N = 2^{\frac{N(N-2)}{4}}  \pi^{-\frac{N}{2}} G^2\left(\frac{N+2}{2} \right)
$ for even $N$ and 
$ 
C_N= 2^{\frac{(N-1)^2}{4}}  \pi^{-\frac{N}{2}} G\left(\frac{N+1}{2} \right)G\left(\frac{N+3}{2} \right) 
$ 
for odd $N$. In these expressions $G(m)$ denotes Barnes' $G$ function \cite{barnes1901}.  

In the presence of stroboscopic measurements with probing time $\tau$, the first detection amplitude $\phi_k$ at time $k\tau$ is constructed from the Loschmidt echo according to Eq.~(2). In particular, for any finite $k$, Eq.~(2) implies the quantum renewal equation for $\phi_k$ \cite{grunbaum2013recurrence,dhar2015b,friedman2016}
\begin{align}
    \phi_k = \cL_k - \sum_{n=1}^{k-1} \cL_{n}\phi_{k-n}.
    \label{appeq:QRE}
\end{align}
Hereafter we adopt the notation $\cL_n = \cL(n \tau)$ for the sake of brevity. The quantum renewal equation \eqref{appeq:QRE} allows us to compute the first detection amplitude $\phi_k$ in the presence of stroboscopic measurements from the sole knowledge of the return amplitude $\cL_k$ in the absence of measurements. 
In particular, the first-detection amplitudes $\phi_k$, with $k\geq 1$, are recursively obtained from the Loschmidt amplitudes. For example, the first four terms read as 
\begin{align}
    \phi_1 &= \cL_1 , \nonumber \\
    \phi_2 &= \cL_2 - \cL_1^2, \nonumber \\
    \phi_3 &= \cL_3 - 2 \cL_1 \cL_2 + \cL_1^3 ,\nonumber \\
    \phi_4 &= \cL_4 -2 \cL_3 \cL_1 - 2 \cL_2^2 + 3 \cL_2 \cL_1^2 - \cL_1^4
\label{appeq:phi4}.
\end{align}
Upon taking the square modulus, one recovers the exact first-detected return probabilities $F_k = |\phi_k|^2$. For larger values of $k$, this recursive approach becomes unfeasible. Accordingly, it is more convenient to consider the 
generating function 
\begin{equation}
\hat{\phi}(z)=\sum_{k=1}^{\infty}\phi_k z^k=\frac{\hat{\cL}(z)}{ 1+ \hat{\cL}(z)}, \quad \mbox{with} \quad \hat{\cL}(z)=\sum_{k=1}^{\infty}\cL_k z^{k},
\label{eq:generating_function_app}
\end{equation}
which is written in terms of the generating function $\hat{\cL}(z)$ of the measurement-free Loschmidt echo. The first detection amplitude $\phi_k$ can now be computed from Eq.~\eqref{eq:generating_function_app} by using a computer algebra system to obtain the symbolic expansion of $\hat{\phi}(z)$ in powers of $z$.
The numerical calculation of $\hat{\cL}(z)$, required to evaluate $\hat{\phi}(z)$, is performed by truncating the sum which defines $\hat{\cL}(z)$ in Eq.~\eqref{eq:generating_function_app} to the first $K$ terms, i.e., $\hat{\cL}(z) \approx \hat{\cL}_K(z)= \sum_{k=1}^{K}\cL_k z^k$. Powers with $k>K$ in $\hat{\cL}(z)$ do not, indeed, contribute to $\phi_K$. The latter is therefore exactly computed from the corresponding series expansion at order $K$ of $\hat{\phi}(z)$ from Eq.~\eqref{eq:generating_function_app}, with $\hat{\cL}(z) \to \hat{\cL}_K(z)$. In this way, we obtain numerically exact data for the first detection probability $F_k=|\phi_k|^2$ as a function of $k$. These data are then compared with the asymptotic predictions presented in the next sections.
The approach described here is computationally feasible on a single computer node for values of $K$ up to $K \approx 500$.

\section{Small-$\tau$ expansion --- the quantum Zeno regime}
\label{sec_app2_Qzeno}
As $\tau \to 0$, the first-detection return probabilities approach $F_{k} = \delta_{k,1}$. This can be  readily seen because the very definition of $\cL(\tau)$ implies that $\lim_{\tau \to 0}\cL(\tau)$ and therefore $\cL_k =1$, for all values of $k$. This, in turn, yields $\lim_{\tau \to 0} \hat{\cL}(z) = z/(1-z)$ [see Eq.~\eqref{eq:generating_function_app}]. 
Inserting the latter expression into Eq.~\eqref{eq:generating_function_app}, one obtains $\hat\phi(z) = z$ and therefore $\phi_k = \delta_{k,1}$. 
This result simply reflects that the target state $\ket{\Psi(0)}$, which coincides with the initial state, is detected with probability 1 at the first measurement. This is  the quantum Zeno effect: under very frequent projective measurements the wavefunction does not evolve from the initial state $\ket{\Psi(0)}$. For small but finite $\tau$, one needs a different approach. As shown in Refs.~\onlinecite{wei2017,krapivsky2018}, the series expansion of $\cL(\tau)$ of $N$ adjacent fermions agrees up to order $\tau^{2N}$ in $\tau$ with the domain wall result of $\cL(\tau) = e^{-\tau^2}$ (the two series expansions, in fact, differ starting from order $\tau^{2N+2}$). This means that, at the leading order in $\tau$, the exact Loschmidt amplitudes given in Eq.~\eqref{appeq:loschmidt_exact} has  the following expansion: $\cL(\tau) = 1 - \tau^2 + \cO(\tau^4)$. Summing over all $\cL_k$, this leads to
\begin{align}
    \hat{\cL}(z) = \sum_{k =1}^{\infty} \cL_k z^k = \frac{z}{1-z} -\tau^2 \Li_{-2}(z) + \cO(\tau^4).
\end{align}
Equation~\eqref{eq:generating_function_app} allows one to obtain $\hat{\phi}(z)$ from $\hat{\cL}(z)$. Expanding $\hat{\phi}(z)$ at the leading order in $\tau^2$, one finds
\begin{align}
    \phi(z) = z - \tau^2 \frac{z+z^2}{1-z} + \cO(\tau^4) = z - \tau^2 (1 + z) \sum_{k=1}^{\infty} z^k=z(1-\tau^2)-2\tau^2\sum_{k=2}^{\infty}z^k,
\end{align}
from which one deduces $\phi_1 = 1 - \tau^2$ and $\phi_{k \geq 2} = -2 \tau^2$. Accordingly
\begin{equation}
F_{k}(N,\tau) = 4 \tau^4+\cO(\tau^6), \quad \mbox{for} \quad k \geq 2, \quad \forall N.
\label{eq:qzeno_FDT_supp}
\end{equation}

\begin{figure}[t]
    \centering
\includegraphics[width=1\textwidth]{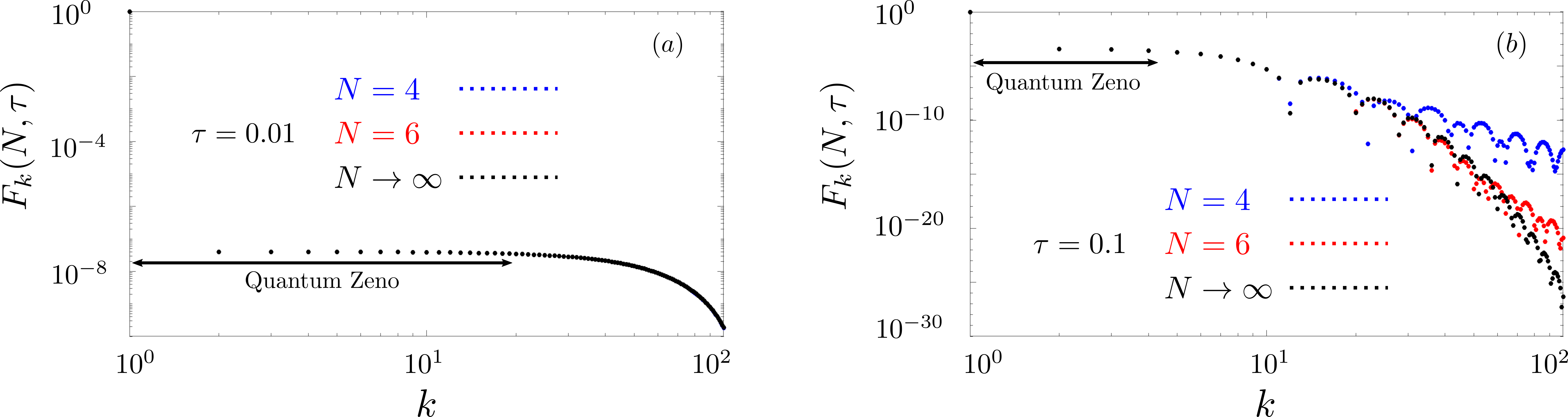}
    \caption{\textbf{Quantum Zeno regime in the many-body FDRT}. (a) Log-log plot of the FDRT probability $F_k(N,\tau)$ as a function of $k$ for fixed value of the stroboscopic time $\tau=0.01$ and for $N=4$, 6, and $\infty$. For $\tau=0.01$, the quantum Zeno approximation in Eq.~\eqref{eq:qzeno_FDT_supp} turns out to be accurate up to $k\lesssim 20$, where the $F_k(N,\tau)$ is constant as a function of $k$. We note that within the quantum Zeno regime the curves corresponding to different finite values of $N$ perfectly coincide with the limit $N\to\infty$. (b) Log-log plot of $F_k(N,\tau)$ as a function of $k$ for a fixed value of $\tau=0.1$ and for the same values of $N$ as in panel (a). In this case, since the value of $\tau$ is larger than that used in panel (a), the range of values of $k$ for which the quantum Zeno prediction in Eq.~\eqref{eq:qzeno_FDT_supp} is accurate shrinks to $k \lesssim 4$. In the Zeno regime, $F_k(N,\tau)$ is again independent of $N$. Upon increasing $k$, instead, $F_k(N,\tau)$ first shows exponential-paramagnetic decay as a function of $k$ according to the $N\to \infty$ curve. Then, the FDRT probability $F_k(N,\tau)$ crosses over to an algebraic decay. The corresponding crossover time, $k_{cr} \approx 30$ for $N=4$ and $k_{cr} \approx 50$ for $N=6$, increases upon increasing $N$. 
    }
\label{fig:qzeno}
\end{figure}

In Fig.~\ref{fig:qzeno}, we check numerically Eq.~\eqref{eq:qzeno_FDT_supp} against the exact evaluation of $F_k$ discussed above. In particular, in panel (a), we plot the FDRT probability $F_k(N,\tau)$ for various (finite) values of $N$ and in the limit $N\to \infty$. We see that for $\tau=0.01$, $F_k(N,\tau)$  
is the largest 
and of order 1 on the first measurement $k=1$. This is expected since $\ket{\Psi(0)}=\ket{\Psi}_T$ and the target state is likely successfully detected at the first attempt for very short probing times $\tau$. For $k\geq 2$, in the Zeno regime, the FDRT probability does not depend on $k$ and it is equal to the value predicted by \eqref{eq:qzeno_FDT_supp}.
The quantum Zeno regime in Fig.~\ref{fig:qzeno}(a) extends up to $k\lesssim 20$ (shown in the figure with the black horizontal line). For larger values of $k$ deviations from the quantum Zeno regime are observed consistently with the fact that the asymptotics \eqref{eq:qzeno_FDT_supp} applies for small values of $k\tau$. Remarkably and unexpectedly, we observe also that the quantum Zeno regime does not depend on the number of fermions $N$. Specifically, the curves for different values of $N$ collapse onto the same curve for $N \to \infty$ (black dots in the figure) within the whole Zeno regime $k\lesssim 20$. In Fig.~\ref{fig:qzeno}(a), moreover, $F_{k=20}(N,0.01) =4 \cdot 10^{-8}$ for $k=20$, which is significantly larger than the value $F_{k=20}(N=6,\tau=2) \sim 10^{-27}$ observed for the same value of $k$ but outside the Zeno regime ($\tau=2$) in Fig.~3(a) of the main text.

In Fig.~\ref{fig:qzeno}(b), we probe deviations from the quantum Zeno regime as $\tau$ increases. In particular, we consider the value $\tau=0.1$ and the same values of $N$ as in Fig.~\ref{fig:qzeno}(a). We see that, correspondingly, 
the range of validity of the quantum Zeno prediction in Eq.~\eqref{eq:qzeno_FDT_supp} for $F_k(N,\tau)$ shrinks to smaller values of $k$. In particular, Eq.~\eqref{eq:qzeno_FDT_supp}  turns out to be accurate only for $k \lesssim 4$. As in panel (a) of this figure, $F_k(N,\tau)$ does not depend on $N$ within the quantum Zeno regime, while it keeps a relatively large value $F_{k=2}(N,0.1)=4\cdot 10^{-4}$. As $k$ increases, deviations from the quantum Zeno prediction are clearly visible. One first observes a fast exponential-paramagnetic decay according to the $N\to \infty$ curve both for $N=4$ and $N=6$. Then, a crossover occurs towards the algebraic-ferromagnetic decay in Eq.~(10) of the main text. The crossover time $k_{cr}$ increases upon increasing $N$, as discussed for Fig.~5 of the main text. Deviations from the quantum Zeno regime, in particular the fast exponential-paramagnetic decay, determine the smaller values of $F_k(N,\tau)$ compared to those in Fig.~\ref{fig:qzeno}(a). 

Summarizing, Fig.~\ref{fig:qzeno} shows that the quantum Zeno regime is observed at small values of $\tau$. The onset of the quantum Zeno regime unexpectedly does not depend on the number of particles $N$. This fact renders the FDRT probability relatively larger for \textit{arbitrary values of} $N$.

\section{Asymptotic behaviour of the first detection return probability of $N$ fermions from branch cut integrals}
\label{app_sec_3_asymptotic}

In this Section, we report the calculation of the large $k$-behaviour of the first detection amplitude $\phi_k$ for $N$ adjacent fermions, reported in Eq.~(10) of the main text. This expression is obtained by evaluating the complex contour integration [see Eq.~\eqref{eq:generating_function_app}]
\begin{equation}
\phi_k = \oint_{C} \frac{\dint z}{2\pi i} \frac{\hat{\phi}(z)}{z^{k+1}},    
\label{supeq:phi_k_complex_int}
\end{equation}
with the generating function $\hat{\phi}(z) = \lim_{w \to -1} \mathcal{Z}_N(z,w)$. 
Similar calculations have been done in Ref.~\onlinecite{friedman2017} in the case of a single particle $N=1$. Accordingly, in this section, we consider a finite $N\geq 2$. As we explain below, the partition function  $\mathcal{Z}_N$ given in Eq.~(9) of the main text can be found from Eq.~\eqref{eq:generating_function_app} and from the asymptotic expression of the Loschmidt echo in Eq.~\eqref{appeq:loschmidt_asymptotic}. In particular, the moment generating function $\hat{\cL}(z)$ of Eq.~\eqref{appeq:loschmidt_asymptotic} reads
\begin{align}
\label{appeq:Lhat_N}
\hat{\cL}(z) &= \sum_{k =1 }^{\infty} \cL_k z^k =  y \Lambda_N(z), 
\end{align}
where $y=e^{-\Delta_N}$, with $\Delta_N = c_N\ln \tau - \ln C_N$. The definition of $\Delta_N$ appears in Eq.~(8) of the main text, which simply follows by taking the logarithm of Eq.~\eqref{appeq:loschmidt_asymptotic} with the replacement $\tau \to \ell \tau$. The constants $c_N$ and $C_N$ have been defined right after Eq.~\eqref{appeq:loschmidt_asymptotic}. The function $\Lambda_N(z)$ is readily identified from the definition of the polylogarithm function: 
\begin{align}
    \Lambda_N(z) =
    \begin{cases} 
        \begin{aligned}
        &\Li_{c_N}(z) & \text{for even $N$},
        \\[2ex]
            & \displaystyle{\frac{e^{-i N \pi/4} \Li_{c_N}(e^{2 i \tau} z) + e^{i N \pi/4} \Li_{c_N}( e^{-2 i \tau} z)}{2}} & \text{for odd $N$ and $\tau \notin \{j \pi/2 \}_{j \in \mathbb{N}}$,} \\[2ex]
           &  \cos \left(N \pi/4 \right) \Li_{c_N}\left((-1)^j z\right)  & \text{for odd $N$ and $\tau = j \frac{\pi}{2}$, with $j \in \mathbb{N}$.}
         \end{aligned}
    \end{cases}
    \label{eq:def_Lambda}
\end{align}
We recall that the polylogarithm function $\Li_s(z)$ is defined as $\Li_s(z) = \sum_{k= 1}^\infty z^k/k^s$ for $|z|<1$, while its definition for $|z|>1$ is obtained by analytic continuation. From Eq.~\eqref{eq:def_Lambda}, we recover the grand-canonical partition function $\mathcal{Z}_N$ given in Eq.~(9) of the main text. This partition function features branch-cut singularities, whose number and position depend on the parity of $N$ and on the value of $\tau$, as we discuss in detail in Subsecs.~\ref{subsec_app_Ngeq4}, \ref{subsec_app_N2}, and \ref{subsec_app_Nodd} below. On the basis of Eq.~\eqref{eq:def_Lambda},  for $N$ odd, we  henceforth refer to the case in which $\tau$ is an integer multiple of $\pi/2$  (bottom line of Eq.~\eqref{eq:def_Lambda}) as a resonant probing time. When, instead, $\tau$ is not an integer multiple of $\pi/2$ (middle line of Eq.~\eqref{eq:def_Lambda}), it will be referred to as non-resonant. The two cases lead to different asymptotics for $\phi_k$.
Here we anticipate the results of this analysis, i.e., the asymptotic decay of $\phi_k$ for large $k$:
\begin{align}
    \phi_k \sim
    \begin{cases}
        \begin{aligned}
            &\frac{y k^{-c_N}}{[1+y \zeta(c_N)]^2} & \text{even $N \geq 4 $}, \\[2ex]
             &\Re \left[\frac{e^{-i(2 k \tau-N \pi/4)}}{(1+y A_{+})^2} \right] y k^{-c_N} & \text{odd $N \geq 3$ and  $\tau \notin \{j \pi/2 \}_{j \in \mathbb{N}}$}, \\[2ex]
             &(-1)^{kj} \frac{\cos\left(N \pi/4 \right)}{\left[1 + y\zeta(c_N) \cos\left(N \pi/4 \right) \right]^2} y k^{-c_N}
             &  \text{odd $N \geq 3$ and $\tau = j \pi/2$ with $j \in \mathbb{N}$,}
        \end{aligned}
    \end{cases}    \label{eq:phi_from_branch_cut}
\end{align}
with $A_{+} = \left[e^{ + i N \pi /4} \zeta(c_N) + e^{- i  N \pi /4} \Li_{c_N}(e^{+ 4 i  \tau}) \right]/2$ and where $\zeta(s) = \sum_{k=1}^{\infty} k^{-s}$ is the Riemann zeta function. Here, for odd $N \geq 3$, the result for resonant values of $\tau = j \pi/2$ equals the limit of the  expression for non-resonant $\tau$ as $\tau \rightarrow j \pi/2$; 
This is in contrast to what happens for $N=1$ \cite{friedman2017}, where the limit of the expression for non-resonant $\tau$ differs from the result corresponding to resonant $\tau$ by a factor of two. This fact is further discussed in Sec.~\ref{subsec_app_Nodd} below. The case $N=2$ shows a rich behavior since the logarithmic corrections to the algebraic decay are present. For this case we perform an asymptotic expansion of $\phi_k$ for small $y$ (large probing time $\tau$) and we find, up to order $y^3$, that
\begin{equation}
\phi_k \sim k^{-1}[B_0(y)+B_1(y)\ln k+B_2(y)\ln^2 k]+\mathcal{O}(y^4), \qquad N=2.
\label{supeq:phi_N2_anticipated_result}
\end{equation}
The expressions of the coefficients $B_{0}(y),B_1(y), B_2(y)$ are reported further below in Eq.~\eqref{supeq:asymptotic_N2_final_2}.
We see that the expansion to order $y^3$ introduces logarithmic corrections up to order $\ln^2 k$ in the asymptotic decay of $\phi_k$ as a function of $k$. The expansion \eqref{supeq:phi_N2_anticipated_result} can be systematically improved by computing higher orders $y^n$ with $n>3$, which produce additional logarithmic corrections up to order $\ln^{n-1} k$.  
One can see that in all the cases $N>2$, up to some possibly oscillating prefactor, the first detection amplitude $\phi_k$ decays always algebraically at large $k$, with $\phi_k \sim k^{-c_N}$. 
The exponent of this decay is the same as that of the Loschmidt echo in Eq.~\eqref{appeq:loschmidt_asymptotic}. 
Also for $N=2$, the exponent $k^{-c_2}=k^{-1}$ of the algebraic decay is set by the Loschmidt echo. In the main text, we show a comparison between the asymptotic prediction in Eq.~\eqref{eq:phi_from_branch_cut} and the exact numerical data (see Sec.~\ref{app_sec_1_exact}) finding excellent agreement. 

In the following Subsections we provide the details of the derivation of Eq.~\eqref{eq:phi_from_branch_cut}. In particular, in Subsec.~\ref{subsec_app_Ngeq4}, we consider the case of even $N\geq 4$. In Subsec.~\ref{subsec_app_N2}, instead, we present the case $N=2$ with the ensuing logarithmic corrections. In Subsec.~\ref{subsec_app_Nodd}, then, we focus on the case with odd $N\geq 3$.

\subsection{Algebraic decay of the first detection return probability for an even number $N\geq 4$ of fermions}
\label{subsec_app_Ngeq4}

\begin{figure}[t]
    \centering
\includegraphics[width=0.4\textwidth]{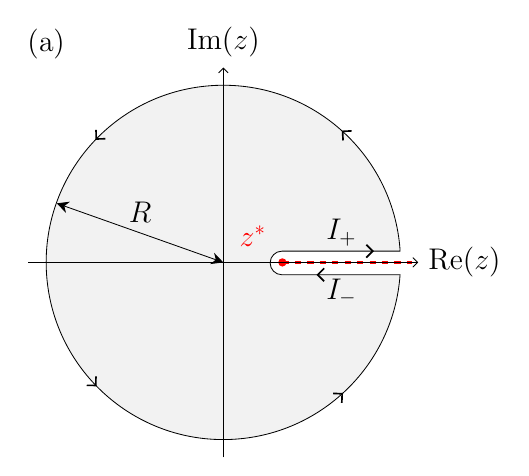}
\includegraphics[width=0.4\textwidth]{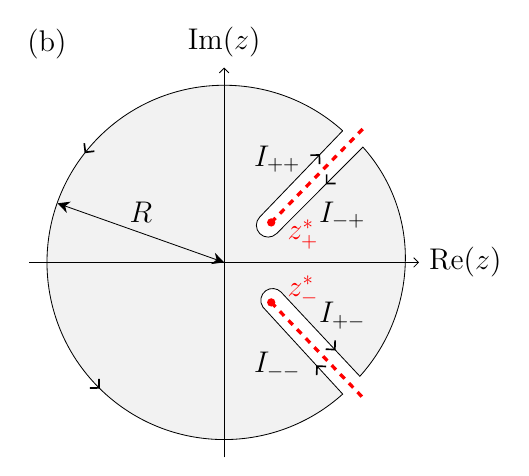}
    \caption{\textbf{Integration path for determining the asymptotic behavior of the first detection amplitude}. In order to evaluate the integral in Eq.~(4), with the generating function $\hat{\phi}(z)$ given in Eq.~(9), we deform the integration contour in the complex plane around the origin. Depending on the parity of $N$, and on the value of $\tau$, $\hat{\phi}(z)$ may display either one or two branch cuts. (a) For even $N$, or odd $N$ but with $\tau = j \pi/2$ for some integer $j$, the function displays a single branch cut on the positive real line starting at $z^* = 1$. (b) For odd $N$ and generic $\tau \notin \{ j\pi/2\}_{j\in\mathbb{N}}$, the function $\hat{\phi}(z)$ displays two branch cuts with branch points at $z^*_{\pm} = e^{\pm 2 i \tau}$, which are depicted in the figure. 
    }
\label{fig:contour_integral}
\end{figure}
We consider here the case in which $N$ is even and larger than or equal to 4. From Eq.~(9) of the main text and Eq.~\eqref{eq:def_Lambda}, the partition function reads $\cZ(z,-1) = y \Li_{c_N}(z) / \left[ 1+ y \Li_{c_N}(z) \right]$. This function displays a single branch cut along $z \in [1,\infty)$, due to the analogous branch cut in $\Li_{c_N}(z)$ on the real axis for $z\in [1,\infty]$ and $c_N \notin \mathbb{Z}$ (in all the calculations, we take the principal value $\mbox{Arg}(z)$ of the argument of a complex number $z$ in the interval $\mbox{Arg}(z)\in (-\pi,\pi]$). At $z=1$, the polylogarithm becomes $\Li_{c_N}(z=1) = \zeta(c_N)$ which is finite because $c_N = N^2/4 >1$. 
The imaginary part of the polylogarithm displays a jump across the cut, with \cite[§1.11.(19)]{bateman1953,NIST:DLMF}
\begin{align}
    {\rm Im}\left[\Li_{c_N}(z \pm i 0^+) \right] = \pm \frac{\pi }{\Gamma(c_N)} \left( \ln z \right)^{c_N-1},
    \label{eq:ImLic}
\end{align}
while the real part is continuous. Here, $\Gamma(c_N)$ denotes the Euler gamma function. When calculating the complex contour integral in Eq.~\eqref{supeq:phi_k_complex_int}, the integration contour 
is deformed in such a way to enclose the branch cut, as shown in Fig.~\ref{fig:contour_integral}$(a)$. Taking the limit of infinite radius $R$ of the contour, while extending the length of the part of the contour which encloses the branch cut at $z \in [1,\infty)$, the only non vanishing contributions to the integral are provided by the line integrals just above and below the branch cut which we denote by
\begin{align}
    I_{\pm} = \int_1^{\infty} \frac{\dint{z}}{2\pi i} \frac{\hat{\phi}(z \pm i 0^+)}{(z\pm i 0^+)^{k+1}}.
\end{align}
The first-detection amplitude in Eq.~\eqref{supeq:phi_k_complex_int} is thus given by
\begin{align}
    \phi_k = I_+ - I_-.
\label{supeq:Ipmbrancheven}
\end{align}
In the subtraction above, all terms which are analytic in $z$ cancel out and one is left with the non-analytic part due to the branch cut, determined by Eq.~\eqref{eq:ImLic}. Since 
\begin{align}
    \hat{\phi}(z) = \frac{\hat{\cL}(z)}{1 + \hat{\cL}(z)} = 1 - \frac{1}{1 + \hat{\cL}(z)} = 1 - \frac{1}{1 + y \Li_{c_N}(z)},
\end{align}
we can write $I_{\pm}$ in Eq.~\eqref{supeq:Ipmbrancheven} as
\begin{align}
I_{\pm}&=\int_{1}^{\infty}\frac{\mbox{d}z}{2\pi i} (z\pm i 0^{+})^{-k-1} \left[1-\frac{1}{1+y \mbox{Li}_{c_N}(z\pm i 0^+)}\right].   
\end{align}
In the large-$k$ limit, the integral is dominated by the region close to $z=1$, i.e., close to the branch-cut singularity of the polylogarithm in $z\in [1,\infty]$. In particular, for $z\to 1^{+}$, the real part of the polylogarithm approaches $\zeta(c_N)$, while the imaginary part is discontinuous across the cut according to Eq.~\eqref{eq:ImLic}. One can then simplify the expression of $I_{\pm}$ by expanding the integral at the leading order around $z=1$ as follows
\begin{align}
I_{\pm}&=\int_{1}^{\infty}\frac{\mbox{d}z}{2\pi i} \mbox{exp}\left[-(k+1)\ln(z\pm i 0^+)\right] \left[1-\frac{1}{1+y \zeta(c_N)\pm iy\pi (\ln z )^{c_N-1}/\Gamma(c_N) }\right] \nonumber \\
&=\int_{1}^{\infty}\frac{\mbox{d}z}{2\pi i} \mbox{exp}\left[-(k+1)\ln z\right] \left[1-\frac{1}{1+y \zeta(c_N)} \pm \frac{iy\pi (\ln z)^{c_N-1}}{\Gamma(c_N)[1+y\zeta(c_N)]^2}+\dots \right],
\end{align}
where $\dots$ denotes subleading orders in the expansion in powers of $(\ln z)^{c_N-1}$ around $z=1$, which we 
neglect for 
 large $k$. We then perform the change of variable $\nu=\ln z$:
\begin{align}
I_{\pm} &=\int_{0}^{\infty}\frac{\mbox{d}\nu}{2\pi i} \mbox{exp}\left(-k \nu \right) \left[1-\frac{1}{1+y \zeta(c_N)} \pm \frac{i y \pi \nu^{c_N-1}}{\Gamma(c_N)[1+y\zeta(c_N)]^2}+\dots \right],\nonumber \\
&=\frac{1}{2\pi i}\left[\frac{1}{k}\left(1-\frac{1}{1+y \zeta(c_N)}\right) \pm \frac{iy \pi k^{-c_N}}{[1+y\zeta(c_N)]^2}+\dots \right].
\label{supeq:Ipm_final}
\end{align}
Inserting the expression in Eq.~\eqref{supeq:Ipm_final} into Eq.~\eqref{supeq:Ipmbrancheven} it is evident that the contribution coming from the analytic terms in the expansion of the polylog around $z=1$ cancel out exactly and only the nonanalytic contribution due to the discontinuity in Eq.~\eqref{eq:ImLic} across the branch cut remains. This leads to the asymptotic expression for the first detection amplitude
\begin{align}
  \phi_k \sim \frac{y k^{-c_N}}{\left[ 1 + y \zeta(c_N)\right]^2} ,
  \label{appeq:phi_k_N_even}
  \end{align}
which neglects subleading terms as $k\to\infty$. 
This is the result anticipated in Eq.~\eqref{eq:phi_from_branch_cut} for even $N\geq 4$. 
For the first detection probability $F_k=|\phi_k|^2$, the exponent $c_N=N^2/2$ of the decay is compared in Fig.~2$(a)$ of the main text with the exact numerical data. The agreement is excellent and it shows that the approximations introduced above are justified and predict the correct exponent of the algebraic decay of $\phi_k$ as a function of $k$. The amplitude of the decay in Eq.~\eqref{appeq:phi_k_N_even}, instead, matches the value computed numerically ans shown in Fig.~2$(b)$ only for large values of $\tau \gtrsim 1.8$. This is caused by the fact that the asymptotics given in Eq.~\eqref{appeq:loschmidt_asymptotic}  (see Ref.~\onlinecite{krapivsky2018}) and used to obtain $\hat{\cL}(z)$, Eqs.~\eqref{appeq:Lhat_N} and \eqref{eq:def_Lambda}, are valid  for $\tau \gg 1$ only.

\subsection{Logarithmic corrections to algebraic decay of first detection return probability of $N=2$ fermions}
\label{subsec_app_N2}

\begin{figure}[t]
    \centering
\includegraphics[width=0.6\columnwidth]{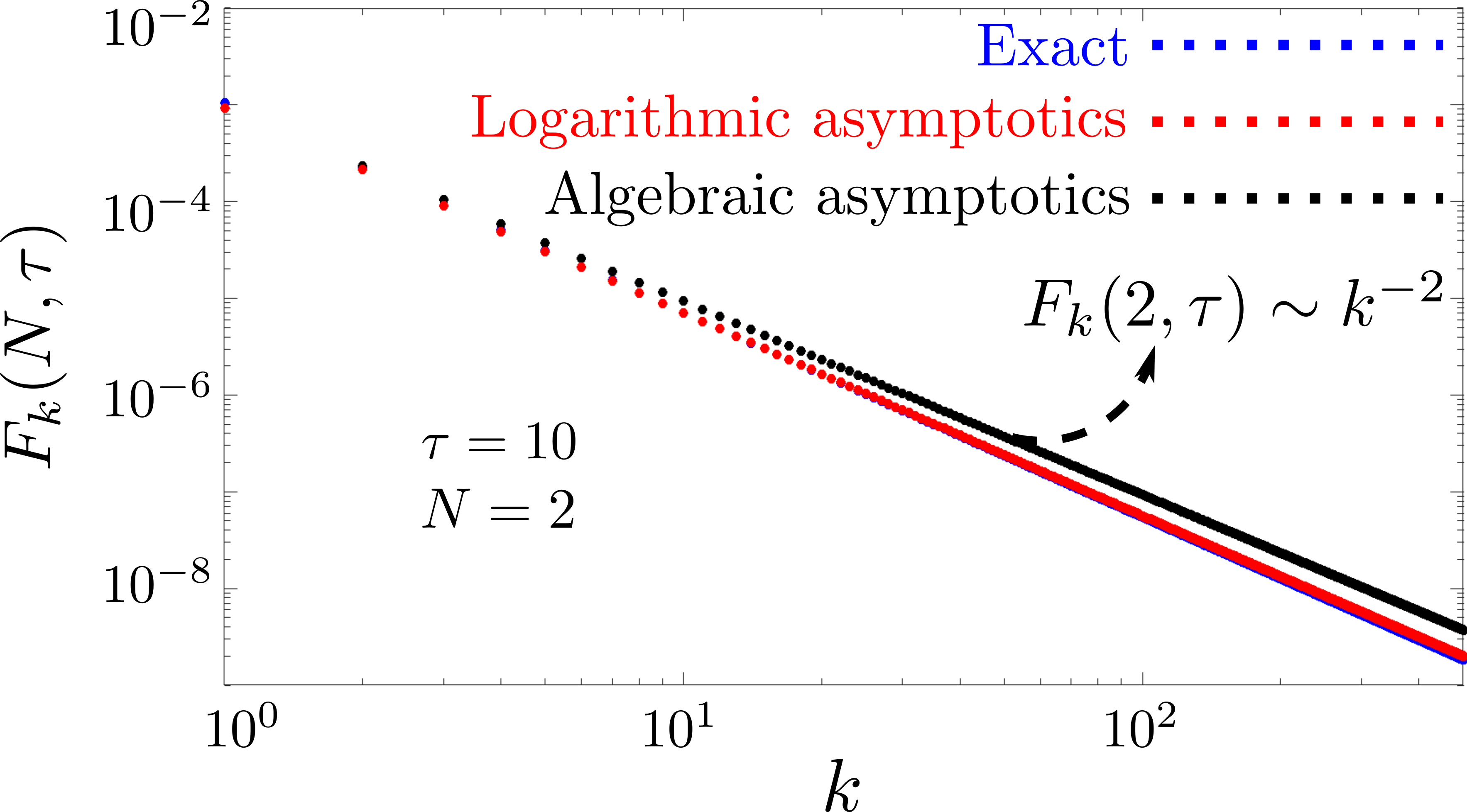}
\caption{\textbf{Ferromagnetic phase - logarithmic corrections to the FDRT probability for $N=2$ fermions}. Log-log plot of the first detection probability $F_k(N,\tau)$ as a function of $k\in [1,500]$ for $N=2$ fermions and probing time $\tau=10$. The blue dotted line is the result of an exact numerical calculation, while the red- and black-dotted lines correspond to different asymptotic expressions. In particular, the black-dotted line is the algebraic asymptotics $F_k=B_0^2(y)/k^2$, with the amplitude $B_0(y)$ given in Eq.~\eqref{supeq:asymptotic_N2_final_2} and setting $B_{1,2}=0$ ($y=C_N \tau^{-c_N}$). The red-dotted line, instead, is the asymptotics in Eq.~\eqref{supeq:asymptotic_N2_final_1}, which account for a logarithmic corrections with coefficients $B_0$ and $B_{1,2}$ from Eq.~\eqref{supeq:asymptotic_N2_final_2}. 
The latter excellently matches the exact data within the range of values of $k$ displayed in the plot. For $k \gtrsim 300$, small discrepancies between Eq.~\eqref{supeq:asymptotic_N2_final_1} and the exact data emerge due to higher-order logarithmic corrections $\ln^{n} k$, with $n\geq 3$.  
}     \label{fig:N2_log_corrections}
\end{figure}

In this Subsection we consider separately the case $N=2$, which is characterized by the fact that the generating function $\phi(z)$ diverges logarithmically at the branch point $z=1$. 
This can be seen from Eqs.~\eqref{appeq:Lhat_N} and \eqref{eq:def_Lambda}, taking into account that  $\Li_1(z) = - \ln(1-z)$. The generating function for $N=2$ fermions is therefore
\begin{align}
    \hat{\phi}(z) = 1 - \frac{1}{1 - y \ln(1-z)}.
\end{align}
As in the case with even $N\geq 4$ discussed in Subsec.~\ref{subsec_app_Ngeq4}, the integral over $\phi(z)$ is performed by deforming the contour of integration into a keyhole contour enclosing the single branch cut located at $z \in [1,\infty)$, as shown in Fig.~\ref{fig:contour_integral}$(a)$. 
The first detection amplitude $\phi_k$ is again written as in Eq.~\eqref{supeq:Ipmbrancheven}, with the integrals $I_{\pm}$ given by 
\begin{align}
I_{\pm}&=\int_{1}^{\infty} \frac{\mbox{d}z}{2\pi i}(z \pm i 0^+)^{-k-1}\left[1-\frac{1}{1-y\ln(1-z\mp i 0^+)} \right] \nonumber \\
&=\int_{1}^{\infty} \frac{\mbox{d}z}{2\pi i}\mbox{exp}[-(k+1)\ln(z\pm i 0^{+})]\left[1-\frac{1}{1-y\ln(1-z\mp i 0^+)} \right] \nonumber \\
&=\int_{1}^{\infty} \frac{\mbox{d}z}{2\pi i}\mbox{exp}[-(k+1)\ln z]\left[1-\frac{1}{1-y\ln(|1-z|)\pm i\pi y} \right]\nonumber \\
&=\int_{0}^{\infty} \frac{\mbox{d}\nu}{2\pi i}e^{-k \nu}\left[1-\frac{1}{1-y\ln(|1-e^{\nu}|)\pm i\pi y} \right] \nonumber \\
&=\int_{0}^{\infty} \frac{\mbox{d}\nu}{2\pi i}e^{-k \nu}\left[1-\frac{1}{1-y\ln \nu \pm i\pi y} +\dots \right]=\frac{1}{2\pi i k}-\int_{0}^{\infty} \frac{\mbox{d}\nu}{2\pi i}e^{-k \nu}\left[\frac{1}{1-y\ln \nu \pm i\pi y} +\dots\right].
\label{supeq:I_pm_N2_intermediate}
\end{align}
In passing from the second to the third equality we used the expression of the branch-cut discontinuity of the logarithm along the negative real axis $\ln(1-z\pm i 0^+)=\ln(|1-z|)\pm i\pi$, while $\ln(z\pm i 0^{+})=\ln z$ since the logarithm is, instead, analytic on the positive real line. In the fourth line we performed the change of variable $z=e^{\nu}$ and we expanded the integrand at the leading order around $\nu=0$, since in the large-$k$ limit, the integral is dominated by the region around $\nu=0$. As in the case $N\geq 4$, in the difference in Eq.~\eqref{supeq:Ipmbrancheven} only terms originating from the branch-cut discontinuity do not cancel out. Accordingly, we eventually obtain the asymptotic expression for $\phi_k$ at large $k$:
\begin{align}
    \phi_k 
    &\sim \int_0^{\infty} \dint{\nu} e^{-k\nu }\frac{y }{\left(1 - y \ln \nu \right)^2 + y^2 \pi^2} .
\label{supeq:N2_final_1}
\end{align}
In order to proceed further from \eqref{supeq:N2_final_1} and determine the logarithmic corrections to the algebraic scaling, we introduce the variable $x=k\nu$:
\begin{equation}
\phi_k \sim \frac{f(\ln k)}{k}, \quad \mbox{with} \quad f(\ln k)= \int_{0}^{\infty} \mbox{d}x \, e^{-x} g_{y,k}(x), \quad \mbox{where} \quad g_{y,k}(x)=\frac{y}{(1-y \ln x +y \ln k)^2 +y^2 \pi^2}.
\label{supeq:N2_final_2}
\end{equation}
The latter equation clearly shows that the FDRT probability $F_k = |\phi_k|^2$ displays, as a function of $k$ and up to logarithmic corrections, an algebraic decay with exponent $2 c_2 =2$, which is dictated by the asymptotic decay of the Loschmidt echo in Eq.~\eqref{appeq:loschmidt_asymptotic}. 
The logarithmic corrections are given by the function $f(\ln k)$, which we evaluate by expanding its integral expression in powers of $y=C_N \tau^{-c_N}$ [see the definition of $y$ after Eq.~\eqref{appeq:Lhat_N}] around $y=0$, i.e., by considering an expansion for long probing time $\tau$. In particular, from Eq.~\eqref{supeq:N2_final_2}, we have that:
\begin{align}
g_{y,k}(x) 
&= y[1-2(y\ln k-y\ln x ) -(y\ln k -y\ln x)^2-y^2\pi^2 +4(y\ln k -y\ln x)^2]+\mathcal{O}(y^4). 
\label{supeq:gfunction_yexpansion}
\end{align}
Integrating this expression term by term, we eventually get the series expansion in $y$ of the function $f(\ln k)$ up to third order in $y$:
\begin{equation}
f(\ln k )=y-2y^2( \ln k +\gamma_{\rm E}) +y^3[3 \ln^2 k +6\gamma_{\rm E}\ln k +3(\gamma_{\rm E}^2+\pi^2/6)-\pi^2]+\mathcal{O}(y^4).
\label{supeq:ffunction_yexpansion}
\end{equation}
In the last equation, we used the mathematical identities
\begin{equation}
\int_{0}^{\infty} \mbox{d}x \,e^{-x}\ln x  =-\gamma_{\rm E}, \quad \mbox{and} \quad \int_{0}^{\infty} \mbox{d}x \, e^{-x} \ln^2 x =\gamma^2_{\rm E}+\pi^2/6,  
\end{equation}
with $\gamma_{\rm E}=0.5772\dots$ the Euler-Mascheroni constant. Plugging Eq.~\eqref{supeq:ffunction_yexpansion} into Eq.~\eqref{supeq:N2_final_2}, we eventually get the following asymptotic behavior for the first detection amplitude $\phi_k$ of $N=2$ fermions:
\begin{equation}
\phi_k \sim k^{-1}[B_0(y)+B_1(y)\ln k +B_2(y)\ln^2 k]+\mathcal{O}(y^4),
\label{supeq:asymptotic_N2_final_1} 
\end{equation}
where the coefficients $B_{0}(y), B_{1}(y), B_{2}(y)$ are given by
\begin{equation}
B_0(y)=y-2y^2\gamma_{\rm E}+3y^{3}(\gamma^2_{\rm E}-\pi^2/6), \quad B_1(y)=-2y^2 + 6 y^3\gamma_{\rm E}, \quad \mbox{and}\quad B_2(y)=3y^3.   
\label{supeq:asymptotic_N2_final_2}
\end{equation}
The expressions in Eqs.~\eqref{supeq:asymptotic_N2_final_1} and \eqref{supeq:asymptotic_N2_final_2} show that the case $N=2$ has a more complex asymptotic behavior than for $N\geq 4$. As a matter of fact, for long probing time $\tau$ (i.e., small $y$) one has $B_0 \sim y$ and $B_{1,2}\sim 0$. This implies that for very large $\tau$, all logarithmic corrections are suppressed and one has $\phi_k \sim \cL_k=y k^{-c_N}$. This is the same behavior as the one observed in Fig.~2$(b)$ of the main text, where the ratio $F_k/\cL_k$ is approximately 1 for large $\tau$. 
For intermediate values of $\tau$  and $N=2$, however, one observes not only a renormalization of the amplitude $B_0$ of the algebraic decay, which leads to $F_k/\cL_k \neq 1$, but also an unexpected appearance of logarithmic corrections $\sim B_{1,2}$. 
In Eqs.~\eqref{supeq:asymptotic_N2_final_1} and \eqref{supeq:asymptotic_N2_final_2}, we see that the expansion up to order $y^3$ generates logarithmic terms up to order $\ln^2 k$. More generically, the order $y^{n+1}$ of the expansion, with $n \geq 1$, is responsible for the emergence of logarithmic terms of order $\ln^n k$. Higher-order logarithmic corrections with $n>2$ become significant at increasingly smaller values of $k$, as $\tau$ decreases (while $y$ increases) correspondingly. 
This is shown in Fig.~\ref{fig:N2_log_corrections} for $\tau=10$, where we compare the exact numerical data for the dependence of $F_k$ on $k$ (see Sec.~\ref{app_sec_1_exact}) with the asymptotic expansion in Eqs.~\eqref{supeq:asymptotic_N2_final_1} and \eqref{supeq:asymptotic_N2_final_2}. One can see that the asymptotic expansion which includes 
logarithmic corrections up to $\ln^2 k$ order excellently describes the exact data in the whole range of $k\in [1,500]$ displayed. In fact, the two curves are barely distinguishable, as small discrepancies emerge only for $k\gtrsim 300$. These discrepancies are caused by higher-order logarithmic corrections and the asymptotic expansion can actually be improved by computing higher-orders terms in the expansion in $y$. 
For the sake of comparison, in Fig.~\ref{fig:N2_log_corrections} we also plot the algebraic asymptotics $F_k \sim B_0^2(y)/k^2$ ($B_{1,2}=0$). The exact data turn out to depart from this algebraic decay already for small values of $k$, showing that logarithmic corrections have to be accounted for in order to properly describe the dependence on $k$ of the first-detection probability $F_k$. 

We note that for $\tau$ values smaller than $\tau=10$ (used in Fig.~\ref{fig:N2_log_corrections}) the asymptotics \eqref{supeq:asymptotic_N2_final_1} and \eqref{supeq:asymptotic_N2_final_2} is not sufficient to describe the data in the $k$ range shown since higher-order logarithmic corrections $\ln^n k$ ($n \geq 3$) are significant. In these cases, however, one can compare the exact data with the complete asymptotics in Eqs.~\eqref{supeq:N2_final_1} and \eqref{supeq:N2_final_2}. This comparison shows that Eqs.~\eqref{supeq:N2_final_1} and \eqref{supeq:N2_final_2} correctly describe the numerically exact data for $F_k$ as functin of $k$ for $\tau$ values up to the order of unity $\tau \gtrsim 1$. For smaller $\tau \lesssim 1$, the exact data show oscillations superimposed to a decay as a function of $k$. We observe that Eqs.~\eqref{supeq:N2_final_1} and \eqref{supeq:N2_final_2} do not capture the oscillations, but they, remarkably, still predict with a good accuracy the decay envelope as a function of $k$. These discrepancies between \eqref{supeq:N2_final_1} and \eqref{supeq:N2_final_2} and the exact data are caused by the fact that the asymptotics \eqref{appeq:loschmidt_asymptotic}, which determines $\Lambda_N$ in Eq.~\eqref{eq:def_Lambda}, does not hold for $\tau \lesssim 1$ (as already commented at the end of Subsec.~\ref{subsec_app_Ngeq4}).

\subsection{Algebraic decay of first detection return probability for an odd number $N \geq 3$ of fermions}
\label{subsec_app_Nodd}

For odd $N \geq 3$, the generating function $\hat{\phi}(z)$ in Eq.~\eqref{eq:generating_function_app} can be calculated via Eqs.~\eqref{appeq:Lhat_N} and \eqref{eq:def_Lambda}. In particular, the function $\Lambda_N(z)$ in the latter displays two branch cuts  along 
\begin{align}
    z_{\pm 1}(u) = u\, e^{\pm 2 i\tau} \quad\mbox{with}\quad u \in [1,\infty),
    \label{eq:z-sigma}
\end{align}
as shown in Fig.~\ref{fig:contour_integral}$(b)$. Importantly, $z_{-1}(1)= z_-^* = e^{-2i\tau}$ is a singular point of the polylogarithm $\mbox{Li}_{c_N}(e^{2i\tau}z)$, while $z_1(1)=z_+^* = e^{2i\tau}$ is a singular point of the polylogarithm $\mbox{Li}_{c_N}(e^{-2i\tau}z)$. 
Note that the two branch cuts above merge into a single branch cut along the real axis whenever the probing time $\tau=j\pi/2$ is resonant (the concept of resonant probing time being defined after Eq.~\eqref{eq:def_Lambda}).
Accordingly, it is convenient to discuss the resonant probing time case separately from the non-resonant one.

\subsubsection{\textbf{Non-resonant time $\tau \notin \{j\pi/2\}_{j\in\mathbb{N}}$}}

\begin{figure}[t]
    \centering
\includegraphics[width=0.6\columnwidth]{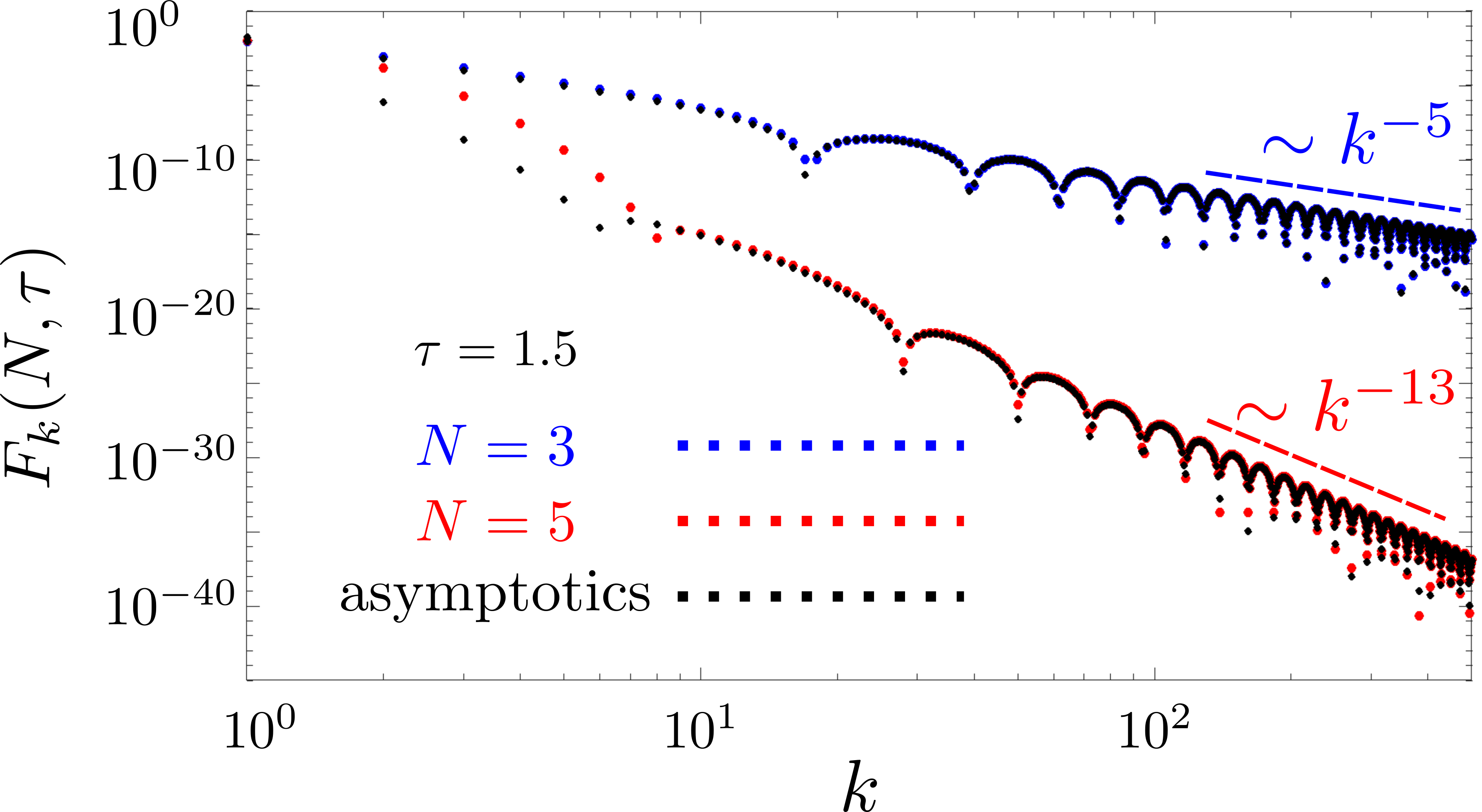}
\caption{\textbf{Ferromagnetic phase - algebraic decay of the FDRT probability with oscillations for odd particle number}. Log-log plot of the first-detection probability $F_k(N,\tau)$ as a function of $k\in [1,500]$ for an odd number $N$ of fermions and $\tau=1.5$. The curves refer to the cases $N=3$ (top-blue dotted line) and $N=5$ (bottom-red dotted line). The first-detection probability $F_k$ decays algebraically as a function of $k$ for large values of $k$, consistently with the mapping to the ferromagnetic phase of the classical spin model. In addition, for odd $N$, persistent oscillations modulate the amplitude of the algebraic decay. The exact numerical data are compared with the asymptotics $F_k=|\phi_k|^2$ in Eq.~\eqref{supeq:nonresonnat_decay} (black-dotted lines for $N=3$ and 5). These analytical expressions correctly predict both the algebraic decay exponent $2c_N$ [see Eq.~\eqref{eq:app-cN}, with $2c_3=5$ and $2c_5=13$] and the amplitude and frequency of the superimposed oscillations.
}     \label{fig:Nodd_plaw_oscillations}
\end{figure}

In order to determine the first-detection amplitudes $\phi_k$ one has to evaluate 
the complex integral given in \eqref{supeq:phi_k_complex_int}. We deform the integration contour around the origin as illustrated in Fig.~\ref{fig:contour_integral}$(b)$, by shaping it into a double keyhole contour enclosing both branch cuts $z_{\pm}(u)$ [see Eq.~\eqref{eq:z-sigma}]. The contribution of the arcs of radius $R$ vanishes as $R \to\infty$ and therefore one is left with four integrals above and below the two branch cuts. Analogously to Ref.~\onlinecite{friedman2017}, where the case $N=1$ is discussed in detail, we introduce
\begin{align}
    I_{\pm,\sigma}  
    &= \frac{1}{2\pi i}\int_1^{\infty} \dint{(e^{2 i \sigma \tau} u)} \left[ \left(u \pm i 0^+\right)e^{2 i \sigma \tau} \right]^{-k-1} \hat{\phi}\left((u  \pm i 0^+) e^{2 i \sigma \tau} \right) \nonumber \\
    &= 
    \frac{e^{-2 i k \sigma \tau}}{2\pi i} \int_1^{\infty} \dint{ u} 
     e^{-(k+1)\ln u}  
    \left\{ 1 - \frac{1}{1+ y\Lambda_N\left(\left(u\pm i 0^+ \right)e^{2 i \sigma \tau}\right)}
    \right\} \nonumber \\
    &= \frac{e^{-2 i k \sigma \tau}}{2\pi i}  \int_0^{\infty} \dint{\nu} e^{-k \nu} \left\{ 1 - \frac{1}{1+ y\Lambda_N\left(\left(e^{\nu} \pm i 0^+ \right)e^{2 i \sigma \tau}\right)}
    \right\},
\label{appeq:ipm_sigma_integral}
\end{align}
where $\sigma = \pm 1$, and where we introduced $\nu = \ln u$, which is analytic along the positive real axis.  
In the notation $I_{\pm,\sigma}$, the subscript $\sigma$ indicates which branch cut \eqref{eq:z-sigma} the integral refers to, while $\pm$ indicates whether the contour is taken outwards ($+$) or inwards ($-$). The four possible terms are shown in Fig.~\ref{fig:contour_integral}$(b)$. By the residue theorem, the integral which determined $\phi_k$ then equals 
\begin{align}
\label{appeq:phi_k_from_4_Is}
    \phi_k = I_{++} - I_{-+} + I_{+-}-I_{--}.
\end{align}
For large $k$, the integrand in Eq.~\eqref{appeq:ipm_sigma_integral} is exponentially suppressed and hence the integral is dominated by the behaviour around $\nu = 0$.
In order to study the resulting leading behaviour, we proceed as follows: for odd $N$, the expression of $c_N$ in Eq.~\eqref{eq:app-cN} never renders an integer, i.e., $c_N \notin \mathbb{N}^+=1,2,3\dots$ and $c_N>1$.
Accordingly, the polylogarithm can be expanded around $\nu = 0$ as (cf. Ref.~\onlinecite{NIST:DLMF}) 
\begin{align}
    \Li_{c_N}(e^{\nu} \pm i 0^+) = \Gamma(1 - c_N) ( - \nu \mp i 0^+)^{c_N - 1} + \sum_{j=0}^{\infty} \frac{\zeta(c_N - j)}{j!} \nu^j.
    \label{eq:polylog_expansion}
\end{align}
The expansion in Eq.~\eqref{eq:polylog_expansion} decomposes into a non-analytic contribution, proportional to $\nu^{c_N - 1}$, and an analytic part, resulting from the analytic continuation in the complex plane of the Riemann $\zeta$ function for $c_N \notin \mathbb{N}$.
Accordingly, $\Lambda_N\left((e^{\nu} \pm i 0^+\right)e^{2 i \sigma \tau})$ (see Eq.~\eqref{eq:def_Lambda}) can be expanded as 
\begin{align}
    \Lambda_N\left((e^{\nu} \pm i 0^+\right)e^{2 i \sigma \tau}) = A_{\sigma} + \frac{e^{i \sigma N \pi/4}}{2}\Gamma(1- c_N)\left(- \nu \mp i 0^+ \right)^{c_N - 1} + \text{analytic terms } \cO(\nu), 
    \label{appeq:lambda_expansion}
\end{align}
where we did not explicitly write down the analytic terms vanishing for $\nu = 0$, and introduced
\begin{align}
\label{appeq:def_Asigma}
    A_{\sigma} = \frac{1}{2} \left[e^{i \sigma N \pi /4} \zeta(c_N) + e^{-i \sigma N \pi /4} \Li_{c_N}(e^{4 i \sigma \tau}) \right].
\end{align}
One then finds that the four integrals $I_{\pm,\sigma}$ have the following expansions around $\nu = 0$:
\begin{align}
\label{appeq:Ipm_expansion}
     I_{\pm,\sigma} &= \frac{e^{-2 i k \sigma \tau}}{2\pi i}  \int_0^{\infty} \dint{\nu} e^{-k \nu} \left[ 1 - \frac{1}{1 + y A_{\sigma}} + e^{i \sigma N \pi/4}\frac{y \Gamma(1-c_N)}{2}\frac{(-\nu \mp i0^+) ^{c_N-1}}{(1 + y A_{\sigma})^2} + \dots  \right] \\
     &= \frac{e^{-2 i k \sigma \tau}}{2\pi i} \left[\left(1 - \frac{1}{1 + y A_{\sigma}} \right) \frac{1}{k} + e^{i \sigma N \pi/4}\frac{y \Gamma(1-c_N)}{2(1+ y A_{\sigma})^2} \int_0^{\infty} \dint{\nu} e^{-k \nu} (-\nu \mp i 0^+)^{c_N - 1} + \dots   \right].
\end{align}
Here, $\dots$ denotes subleading terms which follow both from the analytic terms in $\nu$ from the expansion in Eq.~\eqref{appeq:lambda_expansion} and from higher-order contributions of the expansion of the non-analytic (logarithmic) part around $\nu=0$. All these terms are subleading for large values of $k$. When considering the difference between the integrals done above and below the branch cuts, one finds that the only contributions which do not cancel stem from terms containing non-integer power of $\nu$. 
In fact, following Eq.~\eqref{appeq:phi_k_from_4_Is}, the differences are given by
\begin{align}
    I_{+\sigma} - I_{-\sigma} &= \frac{e^{-2 i k \sigma \tau + i \sigma N \pi/4}}{2\pi i}\frac{y}{2(1+y A_{\sigma})^2} \int_0^{\infty} \dint{\nu} e^{-k \nu} \left( e^{-(c_N-1)\ln(-\nu - i0^+)}  - e^{-(c_N-1)\ln(-\nu + i0^+)}\right) + \ldots \\
    &=\frac{e^{-2 i k \sigma \tau+ i \sigma N \pi/4}}{2\pi i} \frac{y \Gamma(1-c_N)}{2 (1+y A_{\sigma})^2}\left[ e^{- i (c_N - 1) \pi }  - e^{i (c_N - 1) \pi }\right] \int_0^{\infty} \dint{\nu} e^{-k \nu} e^{-(c_N-1) \ln \nu } + \ldots \\
    &=\frac{e^{-2 i k \sigma \tau+ i \sigma N \pi/4}}{2\pi i} \frac{y \Gamma(1-c_N)}{2 (1+y A_{\sigma})^2} 2i  \sin\left(\pi(1-c_N)  \right)    \Gamma(c_N) k^{-c_N} + \ldots \\
    &= \frac{e^{-2 i k \sigma \tau+ i \sigma N \pi/4}}{2(1+y A_{\sigma})^2} y k^{-c_N} + \ldots,
\end{align}
where  the last line is simplified using Euler's reflection formula $\Gamma(z) \Gamma(1-z) = \pi/\sin(\pi z)$ (since $z=1-c_N \notin \mathbb{Z}$). 
Above we have also neglected subleading terms of order $\cO(k^{-c_N -1})$ or higher, which arise when accounting for non-analytic terms of higher-order in $\nu$ in the expansion of the integrand in Eq.~\eqref{appeq:Ipm_expansion}. 
Following Eq.~\eqref{appeq:phi_k_from_4_Is}, the amplitude is then given by
\begin{align}
    \phi_k \sim \frac12 \left[ 
    \frac{e^{-i(2 k \tau-N \pi/4)}}{(1+y A_{+})^2} 
    +
    \frac{e^{i(2 k \tau-N \pi/4)}}{(1+y A_{-})^2}
    \right] y k^{-c_N},
\end{align}
neglecting subleading contributions for large $k$.
Since Eq.~\eqref{appeq:def_Asigma} implies that $A_- = (A_+)^*$, the prefactor of the algebraice decay $\sim k^{-c_N}$ above is indeed real and therefore
\begin{align}
    \phi_k &\sim  \Re \left[
    \frac{e^{-i(2 k \tau-N \pi/4)}}{(1+y A_{+})^2} 
    \right] y k^{-c_N}.
\label{supeq:nonresonnat_decay}
\end{align}
This is the result anticipated in Eq.~\eqref{eq:phi_from_branch_cut}. 
In the case of odd $N$, oscillations are superimposed to the asymptotic algebraic decay. The oscillating prefactor of $F_k$ is, up to a $k$-independent multiplicative constant, determined by the trigonometric functions $\cos^2\left( 2 k \tau - N \pi/4 \right)$, $\sin^2\left( 2 k \tau - N \pi/4 \right)$ 
and $\sin\left( 4 k \tau - N \pi/2 \right)$ and hence oscillates with a period $\pi/(2\tau)$. The function $F_k \, k^{2 c_N}$ is, specifically, periodic for $k\in \mathbb{N}^+$ only if $\tau/\pi$ is a rational number. 
Accordingly, for odd $N$, persistent oscillations are superimposed to the algebraic decay of $\phi_k$ and thus $F_k$. The latter is given by $F_k=|\phi_k|^2 \sim k^{-2 c_N}=k^{-(N^2+1)/2}$ [see Eq.~\eqref{eq:app-cN}], for odd $N\geq 3$, with the exponent $2 c_N$ set by the Loschmidt echo asymptotic decay in Eq.~\eqref{appeq:loschmidt_asymptotic}. 
The case $N=1$, which is discussed in details in Refs.~\onlinecite{friedman2017} with calculations similar to those presented in this section, behaves differently. In fact, the first detection probability decays as $F_k \sim k^{-3}$ (still with oscillations of period $\pi/(2\tau)$), while the  Loschmidt echo decays with an exponent $2 c_1=1$. In Fig.~\ref{fig:Nodd_plaw_oscillations}, we compare the asymptotics in Eq.~\eqref{supeq:nonresonnat_decay} with the results of the exact numerical calculation (see Sec.~\ref{app_sec_1_exact}) and we find excellent agreement as far as the decay exponent $2c_N$ and the amplitude and frequency of the oscillations superimposed to the algebraic decay are concerned.  

\subsubsection{\textbf{Resonant probing time $\tau=j\pi/2$ with $j\in \mathbb{N}$}}  

We consider here the case in which $N$ is odd and $\tau$ is resonant in the sense that $2\tau/\pi$ is an integer $j$.
Then, following Eq.~\eqref{eq:def_Lambda}, $\hat{\cL}(z) = y \cos \left(N \pi/4\right) \Li_{c_N}( (-1)^j z)$ has a single branch cut which lies on either the positive or the negative real axis depending on parity of $j$. 

For even $j$, it follows from \Eqref{eq:def_Lambda} that $\hat{\cL}(z)$ 
for this case (odd $N$ and resonant $\tau$) 
is the same as $\cL(z)$ for even $N$ (in which there is no need to distinguish between resonant and non-resonant $\tau$) upon rescaling $y \mapsto y \cos(N \pi/4)$.
Accordingly, the resulting $\phi_k$ follow directly from Eq.~\eqref{appeq:phi_k_N_even} upon replacing $y$ by $y \cos\left(N \pi/4\right)$.

For odd $j$, instead, Eq.~\eqref{eq:def_Lambda} implies that the generating function $\hat{\cL}(z)$ with odd $N$ (and resonant $\tau$) maps into  $\hat{\cL}(z)$ with even $N$ upon simultaneously replacing $y \mapsto \cos\left(N \pi/4\right) y$ and $z \mapsto -z$. 
Inserting this $\hat{\cL}(z)$ into Eq.~\eqref{eq:generating_function_app}, one finds
\begin{align}
    \phi_k = \frac{1}{2\pi i}\oint \frac{\dint{z}}{z^{k+1}} \frac{\hat{\cL}(z)}{1 + \hat{\cL}(z)} = \frac{(-1)^k}{2\pi i}  \oint \frac{\dint{z}}{z^{k+1}} \frac{\hat{\cL}(-z)}{1 + \hat{\cL}(-z)},
\end{align}
such that, after mapping $z \to -z$, rescaling $y$, and multiplying by $(-1)^k$, one recovers the same integral as the one calculated in Eq.~\eqref{appeq:phi_k_N_even}. 
Exploiting these symmetries, and using the results obtained in \Eqref{appeq:phi_k_N_even}, we therefore obtain for odd $j$
\begin{align}
    \phi_k \sim (-1)^{k}  \frac{\cos\left(N \pi/4 \right)}{\left[1 + y \zeta(c_N) \cos\left(N \pi/4 \right) \right]^2} y k^{-c_N}.
\label{eq:phi_k_N_odd_tau_resonant}
\end{align}
This is the result anticipated in Eq.~\eqref{eq:phi_from_branch_cut}. It shows that also for the case of resonant probing time $\tau$, the first detection probability $F_k \sim k^{-2 c_N}$ decays algebraically with the same exponent as in the non-resonant case in Eq.~\eqref{supeq:nonresonnat_decay}. In the former case, however, the algebraic decay is monotonic and no additional oscillations are present.

In passing, we note that the expression in Eq.~\eqref{eq:phi_k_N_odd_tau_resonant} can be alternatively obtained by 
taking the limit $\tau \to j \pi/2$ of the expression in Eq.~\eqref{supeq:nonresonnat_decay}, corresponding to a non-resonant $\tau$. This is different to the case $N=1$, in which by taking the analogous limit of the expression of $\phi_k$ for non-resonant $\tau$ one overestimates the result with resonant $\tau$ by a factor of two, as discussed in  Ref.~\onlinecite{friedman2017}. The reason for this discrepancy is that, in contrast to $N=1$, none of the two terms in the function \eqref{eq:def_Lambda} diverge at either branch point for $N \geq 3$. Accordingly, the term  $A_{\sigma}$ of $\cO(\nu^0)$ [see Eq.~\eqref{appeq:def_Asigma}] contains contributions from both the terms in \eqref{eq:def_Lambda} (weighted by a half) which, when taking the limit $\tau \to j \pi/2$, correctly recover the contribution of a single branch cut. This is not the case for $N=1$ where $\Lambda_1(z)$ is written in terms of $\Li_{1/2}(z)$ according to Eq.~\eqref{eq:def_Lambda}. The polylogarithm $\Li_{1/2}(z)$ diverges in $z=1$ and therefore the term in Eq.~\eqref{eq:def_Lambda} which remains finite along the branch cut can be neglected as being subleading compared to the other (which is divergent). This causes the constant $A_{\sigma}$ to be determined only by the single term divergent in $z=1$. Consequently, for $N=1$, $\phi_k$ in the nonresonant case turns out to eventually overestimate in the limit $\tau \to j\pi/2$ the resonant first detection amplitude by a factor of $2$. In the present case, $N\geq 3$, this does not take place since $\Li_{c_N}(z)$ is always finite in $z=1$ and the nonresonant result \eqref{eq:phi_k_N_odd_tau_resonant} is obtained continuously as $\tau \to j \pi/2$ in \eqref{supeq:nonresonnat_decay}. 

\section{Effective exponent and the onset of algebraic behavior}
\label{sec:effective_exp}
\begin{figure}[t]
    \centering
\includegraphics[width=0.8\linewidth]{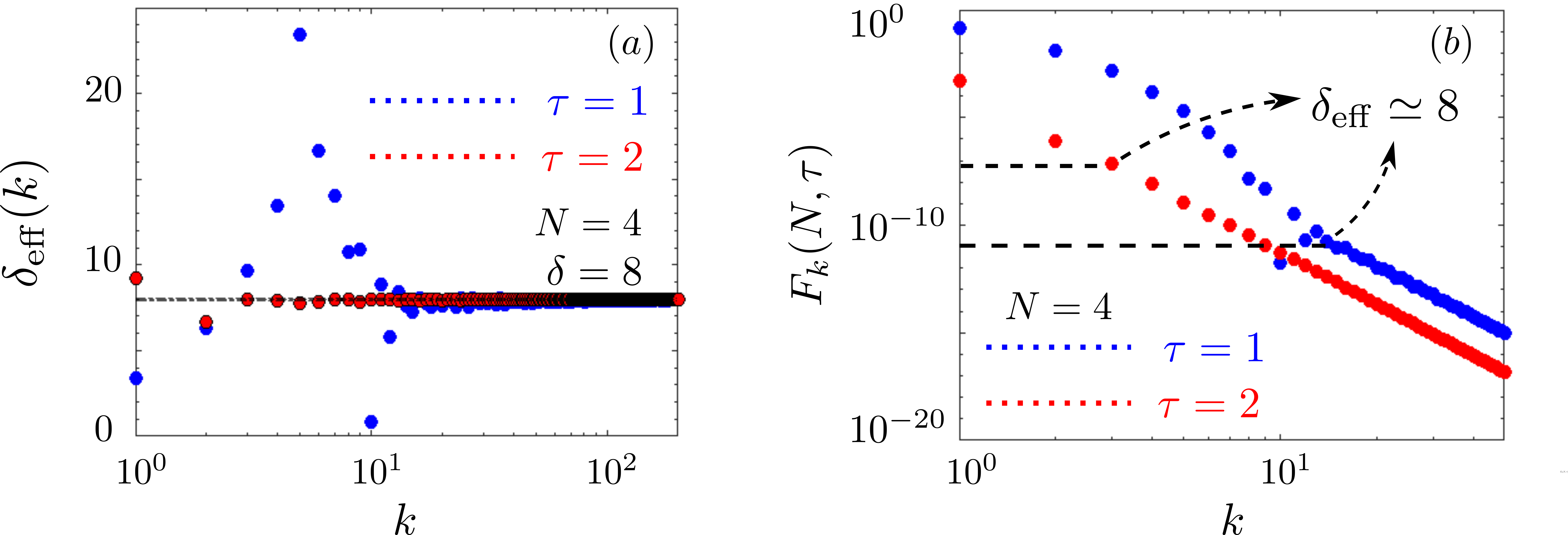}
\caption{\textbf{Effective decay exponent for different probing times}. (a) Linear-log plot of the decay exponent $\delta_{\mathrm{eff}}(k)$ in Eq.~\eqref{supeq:effective_exp} as a function of $k$ for $N=4$ particles and two different probing time values $\tau=1$ and $\tau=2$. We see that $k_{cr}(N,\tau)$ decreases and the convergence of $\delta_{\mathrm{eff}}(k)$ to the actual decay exponent $\delta=8$ is accordingly accelerated as $\tau$ is increased. In particular, for $\tau=2$ one has $\delta_{\mathrm{eff}} \approx 8$ for $k \gtrsim k_{cr}$, with $k_{cr}(N=4,\tau=2)\approx 3$. (b) Log-log plot of the FDRT probability $F_k(N,\tau)$ as a function of $k$ for the same values of $N$ and $\tau=1$ (blue-dotted top line) and $\tau=2$ (red-dotted bottom line) as in panel (a). The black horizontal dashed lines mark the values of $F_k(N,\tau)$ corresponding to convergence of $\delta_{\mathrm{eff}}(k) \approx \delta$ fr $k \gtrsim k_{cr}$. Namely, $F_{k=3}(N=4,\tau=2) \sim 7.5 \cdot 10^{-8}$ (top dashed line) for $k_{cr}(N=4,\tau=2) \approx 3$ and $F_{k=15}(N=4,\tau=1) \sim 10^{-11}$ for $k_{cr}(N=4,\tau=1) \approx 15$ (bottom dashed line).}
\label{fig:effective_exps}
\end{figure}

In this Section, we investigate in more detail the onset of the algebraic-ferromagnetic decay as a function of $k$. In particular, we focus on the value $k_{cr}$ beyond which the FDRT probability $F_k(N,\tau)$ shows an algebraic dependence on $k$ and show that $k_{cr}$ can be reduced 
upon increasing the probing time $\tau$. This allows one to detect the algebraic-ferromagnetic decay with few measurements at early times, when the FDRT probability $F_k(N,\tau)$ is still relatively large.

In order to quantify the onset of an algebraic decay, we focus on the effective exponent
\begin{equation}
\delta_{\mathrm{eff}}(k)=-\frac{\log (F_{b k}(N,\tau)/F_{k}(N,\tau))}{\log b},
\label{supeq:effective_exp}
\end{equation}
with $b>0$ a scaling factor, which we fix to $b=2$.
In fact, for an asymptotic behavior $F_k(N,\tau) \sim k^{-\delta}$, this effective exponent $\delta_{\mathrm{eff}}(k)$ 
converges to the exponent $\delta$ for $k\gtrsim k_{cr}$.

Figure~\ref{fig:effective_exps} shows the effective exponent $\delta_{\mathrm{eff}}(k)$ and the algebraic-ferromagnetic decay of the probability of the FDRT for various values of $\tau$ and fixed particle number $N=4$. In particular, in panel (a), we observe that $\delta_{\mathrm{eff}}(k)$ attains the expected value $\delta=2c_4=8$ of the algebraic decay at increasingly smaller values of $k$ as the probing time $\tau$ increases. The crossover time $k_{cr}(N,\tau)$ accordingly decreases as $\tau$ is increased for a fixed number $N$ of fermions. Namely, for $\tau=1$, the effective exponent $\delta_{\mathrm{eff}}(k) \approx \delta$ for $k\gtrsim k_{cr}(N=4,\tau=1)$, with $k_{cr}(N=4,\tau=1)\approx 15$. For $\tau=2$, instead, the convergence to $\delta$ occurs already  for $k \gtrsim k_{cr}(N=4,\tau=2)$, with $k_{cr}(N=4,\tau=2)\approx 3$. This observation shows that in order to observe the asymptotic algebraic decay of $F_k(N,\tau)$ as a function of $k$ it is \emph{not} necessary to perform a large number $k$ of measurements. This is crucial, because $F_k$ rapidly becomes practically undetectable upon increasing $k$, as shown in Fig.~3 of the main text, where $F_k \sim 1.5 \cdot 10^{-17}$ at $k=50$. Such a small value of the FDRT probability is certainly too small to be measurable. 
However, exploiting the low value of $k_{cr}(N,\tau)$ as $\tau$ is increased and the consequent fast convergence of $\delta_{\mathrm{eff}}$ in Fig.~\ref{fig:effective_exps}(a), one can tune the onset of the algebraic decay to smaller values of $k$, at which the probability $F_k(N,\tau)$ is relatively large and therefore more accessible. For example, in Fig.~\ref{fig:effective_exps}(b) the black horizontal dashed lines correspond, from bottom to top, to $F_{k=k_{cr}}(N=4,\tau=1) \simeq 10^{-11}$ for $k_{cr}(N=4,\tau=1)\approx 15$ and $F_{k=k_{cr}}(N=4,\tau=2) \simeq 7.5 \cdot 10^{-8}$ for $k_{cr}(N=4,\tau=2)\approx 3$.

We eventually comment here on the experimental accessibility of the FDRT probability based on the findings above. A promising experimental platform to measure $F_k$ is arguably provided by quantum computers, as recently shown in Refs.~\onlinecite{tornow2023,wang2024first,yin2024restart}. In this platform, stroboscopic measurements can be conveniently implemented via midcircuit readouts with up to $k \leq 40$ midcircuit readouts. These values of $k$ are certainly compatible with the low number of measurements necessary to observe the algebraic decay in Fig.~\ref{fig:effective_exps}(b) (for instance $k\sim k_{cr}(N=4,\tau=2) \sim 3,4$ measurements are required for $\tau=2$). Furthermore, as commented above, the FDRT probability is mostly concentrated on the first few measurements, after which it becomes rapidly small. Readout errors in first few measurements can be successfully reduced via error-mitigation techniques \cite{tornow2023}, which allow to precisely measure $F_k$ for the first values of $k$. We therefore envision that the predictions in Fig.~\ref{fig:effective_exps}(b) are compatible with the time scales that are within reach of current experiments. It is, however, important to note that a limiting factor for the observation of our findings on quantum computers is given by the spatial scales. The analysis of Refs.~\onlinecite{tornow2023,wang2024first,yin2024restart} specifically concerns a single qbit (i..e, the case $N=1$) on rings of finite lengths $L=2$ and $3$. For such sizes in Refs.~\onlinecite{tornow2023,wang2024first,yin2024restart} a large number of $32000$ runs of the experiment on the quantum computer is performed, which further renders the estimate of $F_k$ at short times $k$ more accurate. Our predictions, instead, apply for larger systems on a ring of infinite length $L\to \infty$, as written after Eq.~\eqref{eq:free_fermions_Hamiltonian}. Our analysis of Sec.~\ref{app_sec_1_exact} can be, however, simply extended to the case of a chain of finite length $L$. In this way we expect to be able to directly connect the present analysis to modern experiments on quantum computer platforms.

\section{Fa\`{a} di Bruno approximation for the FDRT probability for large $\tau$ and $N=\infty$}
\label{app_sec_4_Bruno}

The generating function $\hat{\phi}(z)$ for the first detection amplitudes $\{\phi_k\}_k$ in Eq.~(11) of the main text  (see also Eqs.~\eqref{eq:generating_function_app} 
can be written as the composition $f(g(z))$ with $f(z) = z/(1+z)$ and $g(z) = \hat{\cL}(z) = \Theta_+(z,q=e^{-\tau^2}) =\sum_{k=1}^{\infty} z^k q^{k^2}$.   
Applying the Fa\`{a} di Bruno formula for the derivative of a composite function (see, e.g., Ref.~\onlinecite{comtet1974}), we can evaluate the $\cO(z^k)$ expansion of $\hat{\phi}(z) = f(g(z))$, finding
\begin{align}
    \phi_k = \frac{1}{k!}\left.\frac{\mathrm{d}^k}{\mathrm{d} z^k} f(g(z)) \right|_{z=0}
    = \sum_{m_1,...,m_k} \frac{1}{m_1! (1!)^{m_1}\cdot \ldots \cdot m_k! (k!)^{m_k}} f^{(m_1 + \ldots + m_k)}(g(0)) \prod_{j=1}^{k} \left( g^{(j)}(0) \right)^{m_j},
\end{align}
where the sum runs over all partitions $\left\lbrace m_j  \right \rbrace > 0$ of non-negative integers satisfying $m_1 + 2 m_2 + \ldots + k m_k = k$. This sum is simplified by observing that $g(0) = 0$, $g^{(j)}(0) = j! q^{j^2}$, where  $q=e^{-\tau^2}$, and $f^{(n)}(g(0)) = f^{(n)}(0) = (-1)^{n-1} n!$, leading to
\begin{align}
   \phi_k &=  -\sum_{m_1,...,m_k} \frac{1}{\prod_{j'=1}^k m_{j'}! (j'!)^{m_{j'}} } (-1)^{m_1 + ... + m_k} (m_1 + ... + m_k)! \prod_{j = 1}^k \left( j! q^{j^2} \right)^{m_j} \nonumber \\
   &=-\sum_{m_1,...,m_k} (-1)^{m_1 + ... + m_k} \frac{(m_1 + ... + m_k)!}{m_1! \ldots m_k!} \prod_{j = 1}^k \left( q^{j^2} \right)^{m_j}.
   \label{appeq:phi_k_FdB_sum}
\end{align}
Each term in the previous sum can be interpreted in terms of the spin model, as discussed in the main text. In fact, one can identify $m_j$ as the number of domains of length $j$ and thus $r = \sum_{j} m_j$ as the total number of domains in the chain, which has a total volume $k = \sum_j j m_j$. As it can be seen from Eq.~\eqref{appeq:phi_k_FdB_sum}, the leading order contribution to $\phi_k$ is of perturbative order $\cO(q^k)$ and arises from the complete partition of the system into domains of length one, corresponding to having $m_1 = k$ and $m_{j} = 0$ for $j \geq 2$. The exact expression in Eq.~\eqref{appeq:phi_k_FdB_sum} can then be used to determine the subleading corrections to term of order $q^k$, due to domains of length larger than one. Namely, we perform a perturbative expansion as follows. 

We develop a systematic expansion of $\phi_k$ in $q$ by converting the exact expression 
\eqref{appeq:phi_k_FdB_sum} into a perturbative expansion by truncating the power series of $\hat{\cL}(z)$ at some finite power $K$, leading to $g(z) = qz + q^4 z^2 + ... + q^{K^2} z^K$. This means that the summation in Eq.~\eqref{appeq:phi_k_FdB_sum}, is restricted to partitions $\{ m_j\}$ with $m_{j>K} = 0$. 
Physically, this corresponds to considering a spin model in which only domains of length up to $K$ are allowed. For $K\geq k$, this restriction does not introduce any error in the computation of $\phi_k$ as no domain can be larger than the volume $k$. 
For $K\leq k-1$, instead, the truncation affects the value of the sum. The associated truncation error 
is determined by the lowest-order contribution with a single domain of length $K+1$ and then $k-(K+1)$ domains of length $1$. Following Eq.~\eqref{appeq:phi_k_FdB_sum}, such a configuration is obtained by setting $m_1 = k- (K+1), m_2 = ... = m_{K} = 0, m_{K+1} = 1$, and is of perturbative order $\cO(q^{K^2+K+k})$.
This means that, after dividing by the leading $q^k$ term, the perturbative error obtained from truncating the sum at the domain length $K$ is given by $\phi_k/q^k \sim \cO(q^{K^2 + K})$.

The first non-trivial approximation is for $K=2$, where
\begin{align}
    \phi_k = -\sum_{m_2=0}^{\floor{k/2}} (-1)^{k-m_2} \frac{(k-m_2)!}{(k-2m_2)! m_2!} q^{k+2m_2} = (-1)^{k-1} q^k \left[ \sum_{m_2=0}^{\floor{k/2}} {k-m_2 \choose m_2} \left(-q^2\right)^{m_2} + \mathcal{O}\left(q^6\right) \right].
\end{align}
Hereafter, we denote by $\floor{x}$ the greatest integer smaller or equal than $x$. A serendipitous identity, provided in Eq.~(1.60) of Ref.~\onlinecite{gould1972}, is
\begin{align}
    \sum_{m=0}^{\floor{k/2}} {k-m\choose m} (-1)^m\left(u v \right)^{m} (u+v)^{k-2m} = \frac{u^{k+1}-v^{k+1}}{u-v}.
\label{eq:gould_identity}
\end{align}
Setting 
\begin{align}
    u = \frac{1+\sqrt{1- 4 q^2}}{2}, \qquad v = \frac{1-\sqrt{1- 4 q^2}}{2},
\end{align}
one obtains from Eq.~\eqref{eq:gould_identity} the $K=2$ Fa\`a-di-Bruno approximation ($q=e^{-\tau^2}$)
\begin{align}
    \phi_k = (-1)^{k-1}e^{-k \tau^2} \left[\frac{\left( 1 + \sqrt{1-4e^{-2 \tau^2}}\right)^{k+1} - \left( 1 - \sqrt{1-4e^{-2 \tau^2}}\right)^{k+1}}{2^{k+1}\sqrt{1-4e^{-2 \tau^2}}} + \cO(e^{-6 \tau^2}) \right].
\label{supeq:K2Bruno}
\end{align}
For $\tau \gtrsim 1$, this expression correctly reproduces the limiting behaviour of $\phi_k = (- e^{- \tau^2})^k$ for large $k$. Moreover, it also correctly predicts  the transition from a monotonic 
to an oscillatory behavior as a function of $\tau$, shown in Fig.~3$(b)$ of the main text.  

An even better asymptotics is obtained by choosing $K =3$, where the Fa\`{a}-di-Bruno approximation explicitly reads
\begin{align}
    \phi_k = (-1)^{k-1}q^{k} \left[ \sum_{m_2=0}^{\floor{k/2}} (-1)^{m_2} \sum_{m_3 = 0}^{\floor{(k-2m_2)/3)}}\frac{(k-m_2-2m_3)!}{(k-2m_2-3m_3)! m_2! m_3!} q^{2m_2 + 6m_3} + \cO(e^{-12\tau^2}) \right]. 
\label{eq:bruno_third_order}     
\end{align}
We were not able to find an elementary analytic expression of this double sum. We, however, calculated numerically Eq.~\eqref{eq:bruno_third_order}, which has been used to produce the asymptotic results reported as black dashed lines in Fig.~$3(b)$ of the main text. From this figure, we see that Eq.~\eqref{eq:bruno_third_order} already excellently captures the dependence of $F_k(\tau)$ for $\tau>\tau_c$ (with $\tau_c=0.77\dots$ defined after Eq.~(11) of the main text). At $\tau_c$, however, the previously determined asymptotic behavior departs from the numerical data and a sharp transition occurs towards an oscillatory behaviour. This corresponds to the onset of imaginary roots in the partial theta function, as shown also in Fig.~3$(a)$. In particular, Fig.~3$(b)$ shows that $F_k(\tau)$ vanishes at $k-2$ special values of $\tau$ within the interval $(0,\tau_c)$. These roots can be physically understood in terms of the entropically favourable formation of up to $r=1,2\dots k-1$ longer domains as $\tau$ decreases. Due to the quantum nature of $Z_\infty(k,-1)$ in Eq.~(6), which requires $w\to -1$, the associated amplitudes interfere destructively, leading to a vanishing $F_k=|Z_{\infty}(k,-1)|^2$.
Mathematically, these roots can also be understood as follows: expanding Eq.~(5) in powers of $q=e^{-\tau^2}$ results in a polynomial with alternating signs and no constant term for  which  Descartes' rule of signs states that the number of positive real roots is at most $k-2$. We have done an extensive numerical search of these roots for various values of $k$, finding that this bound is actually always saturated. However, we cannot currently provide a  proof of this fact.

\end{document}

%% file: macros.tex
\usepackage{xcolor}
\usepackage{url}

\newcommand{\dint}[1]{{\rm d}{#1}\,}

\renewcommand{\Re}{\mathrm{Re}}

\def\adj{\dagger}

\newcommand{\Li}{\operatorname{Li}}

\def\bJ{\mathbf{J}}

\def\cH{\mathcal{H}}

\def\cL{\mathcal{L}}

\def\cO{\mathcal{O}}

\def\cZ{\mathcal{Z}}

\newcommand{\Eqref}[1]{Eq.~\eqref{#1}}

\newcommand{\ket}[1]{\left| #1 \right \rangle}
\newcommand{\bra}[1]{\left\langle #1 \right|}
\newcommand{\braket}[2]{\left\langle #1 | #2 \right \rangle}